\documentclass[aip]{revtex4-1}

\usepackage[T1]{fontenc}
\usepackage[utf8]{inputenc}
\usepackage{amsmath}
\usepackage{amsthm}
\usepackage{amssymb}
\usepackage{graphicx}
\usepackage{xargs}[2008/03/08]
\usepackage[unicode=true,
 bookmarks=false,
 breaklinks=false,pdfborder={0 0 1},backref=false,colorlinks=false]
 {hyperref}
\hypersetup{pdftitle={A continuous model for CISS},
  pdfauthor={Matthias Geyer}}
  
\makeatletter

\usepackage{float}
\usepackage{braket}
\usepackage{tikz-cd}

\draft 

\begin{document}

\newcommand\uptextrm[1]{\textrm{\textup{#1}}}

\newcommand\bbC{\mathbb{C}}
\newcommand\bbR{\mathbb{R}}
\newcommand\bbN{\mathbb{N}}
\newcommand\bbZ{\mathbb{Z}}
\newcommand\bbH{\mathbb{H}}

\newcommand\calH{\mathcal{H}}
\newcommand\calL{\mathcal{L}}
\newcommand\calF{\mathcal{F}}
\newcommand\calS{\mathcal{S}}
\newcommand\calB{\mathcal{B}}

\newcommand\calE{\mathcal{E}}

\newcommand\der{\uptextrm d}

\newcommandx\partder[2][usedefault, addprefix=\global, 1=]{\frac{\partial#1}{\partial#2}}

\newcommandx\partdern[3][usedefault, addprefix=\global, 1=]{\frac{\partial^{#3}#1}{\partial#2^{#3}}}

\newcommand\setin{\subseteq}

\newcommand\Ln{\operatorname{Ln}}
\newcommand\Arg{\operatorname{Arg} }
\newcommand\img{\operatorname{im}}
\newcommand\sign{\operatorname{sgn}}
\newcommand\inter{\operatorname{int}}
\newcommand\trace{\operatorname{tr}}
\newcommand\Res{\operatorname{Res}}
\newcommand\ind{\operatorname{ind}}
\newcommand\Div{\operatorname{div}}
\newcommand\Grad{\operatorname{grad}}
\newcommand\lin{\operatorname{lin}}
\newcommand\ran{\operatorname{ran}}
\newcommand\dist{\operatorname{dist}}
\newcommand\ident{\operatorname{id}}

\newcommand\markequal[1]{\operatorname*{=}^{#1}}

\newcommand\subrarrow[1]{\operatorname*{\longrightarrow}_{#1}}
\newcommand\superlrarrow[1]{\operatorname*{\longrightarrow}^{#1}}

\newcommand\fun[3]{#1\colon#2\rightarrow#3}

\newcommandx\integ[4][usedefault, addprefix=\global, 1=, 2=]{\intop_{#1}^{#2}\!#3\,\uptextrm d#4}
\newcommandx\integsht[3][usedefault, addprefix=\global, 1=, 2=]{\intop_{#1}^{#2}\!#3}

\newcommand\setm[2]{\left\{  #1\left|#2\right.\right\}  }

\newcommand\restr[2]{\left.#1\right|_{#2}}

\newcommand\comp[1]{#1^{\uptextrm C}}

\newcommand\Norm[1]{\left\Vert #1\right\Vert }

\newcommand\supp{\operatorname{supp}}
\newcommand\essup{\operatorname*{ess\, sup}}
\newcommand\essran{\operatorname{ess\, ran}}

\newcommand\LinHp{\mathcal{L}\left(\calH^{\prime}\right)}
\newcommand\vphi{\varphi}
\newcommand\veps{\varepsilon}

\newcommand\Hedi{H_{\textrm{diag}}^{\veps}}
\newcommand\bigO{\mathcal{O}}
\newcommand\Hprime{\calH^{\prime}}

\newcommand\heberg[1]{#1_{\uptextrm H}}

\newcommand\ccinfty{C_{\uptextrm c}^{\infty}}

\newcommand\cosphi[1]{\cos\phi_{#1}}
\newcommand\sinphi[1]{\sin\phi_{#1}}

\newcommand\doublesum[3]{\sum_{#1}^{\begin{array}{c}
 {\scriptstyle #2}\\
 {\scriptstyle #3} 
\end{array}}}

\newcommand\lamdsig[1]{\boldsymbol{\lambda}_{#1}\cdot\boldsymbol{\sigma}}
\newcommand\lamdsigs[2]{\boldsymbol{\lambda}_{#1}\left(#2\right)\cdot\boldsymbol{\sigma}}

\newcommand\Heff{H_{\uptextrm{eff}}}
\newcommand\Hkine{H_{\uptextrm{kin},\bbR^{3}}}
\newcommand\Hsoce{H_{\uptextrm{SOC},\bbR^{3}}}
\newcommand\Hrash{H_{\uptextrm{Rashba}}}
\newcommand\Heuc{H_{\bbR^{3}}}
\newcommand\Hcurv{H_{\textrm{curv}}}
\newcommand\Heuceps{H_{\bbR^{3}}^{\veps}}
\newcommand\Hcurveps{H_{\textrm{curv}}^{\veps}}

\newcommand\Hkinc{H_{\uptextrm{kin},\uptextrm c}}
\newcommand\Hsocc{H_{\uptextrm{SOC},\uptextrm c}}
\newcommand\Hkinceps{H_{\uptextrm{kin},\uptextrm c}^{\veps}}
\newcommand\Hsocceps{H_{\uptextrm{SOC},\uptextrm c}^{\veps}}
\newcommand\Hkineps{H_{\textrm{kin}}^{\veps}}
\newcommand\Hsoceps{H_{\textrm{SOC}}^{\veps}}
\newcommand\Hefftilde{\widetilde{H}_{\textrm{eff}}}

\newcommand\mmcom{\textrm{,}}
\newcommand\mmdot{\textrm{.}}

\newcommand\normtilde[1]{\tilde{\left|#1\right|}}
\newcommand\wtilde[1]{\widetilde{#1}}

\newcommandx\adj[2][usedefault, addprefix=\global, 1=]{#2_{#1}^{\dagger}}


\title{Effective Hamiltonian model for helically constrained quantum systems within adiabatic perturbation theory: application to the Chirality-Induced Spin Selectivity (CISS) Effect} 



\author{Matthias Geyer}
\email[e-mail: ]{matthias.geyer@tu-dresden.de}
\altaffiliation{Dresden Center for Computational Materials Science (DCMS), TU Dresden, 01062 Dresden, Germany}
\author{Rafael Gutierrez}
\email[e-mail: ]{rafael.gutierrez@tu-dresden.de}
\author{Gianaurelio Cuniberti}
\email[e-mail: ]{gianaurelio.cuniberti@tu-dresden.de}
\altaffiliation{Dresden Center for Computational Materials Science (DCMS), TU Dresden, 01062 Dresden, Germany}
\affiliation{Institute for Materials Science and Max Bergmann Center of Biomaterials, TU Dresden, 01062 Dresden, Germany}

\makeatother

\date{\today}

\begin{abstract}
The chirality-induced spin selectivity (CISS) effect has been confirmed 
experimentally for a large class of organic molecules. Adequately modeling 
the effect remains a challenging task, with both phenomenological models 
and first-principle simulations yielding inconclusive results. Building upon a previously presented model by K. Michaeli and R. Naaman (J. Phys. Chem C 123, 17043 (2019)) we systematically investigate an 
effective 1-dimensional model derived as the limit of a 
3-dimensional quantum system with strong  confinement and including
spin-orbit coupling. Having a simple analytic structure, such models 
can be considered a minimal setup for the description of spin-dependent effects.
We use adiabatic perturbation theory to provide a mathematically sound 
approximation procedure applicable to a large class of spin-dependent 
continuum models. We take advantage of the models simplicity by 
analyzing its structure to gain a better understanding how the occurrence and 
magnitude of spin polarization effects relate to the model's parameters and geometry.
\end{abstract}


\maketitle 


\section{Introduction}

Spin-selective transfer processes related to chiral symmetries of
molecular systems have increasingly attracted the attention over the
past few years.\cite{Mishra14872,doi:10.1063/1.4966237,doi:10.1021/acsnano.5b00832,ADMA:ADMA201504725,C6CS00369A,SMLL:SMLL201602519,Kumar2017,doi:10.1021/jp509974z,doi:10.1146/annurev-physchem-040214-121554,doi:10.1021/jacs.6b12971,BenDor2017,doi:10.1021/jacs.6b10538,doi:10.1021/acs.jpclett.8b00208,doi:10.1021/jacs.8b08421}
The effect, which has been termed Chirality-Induced Spin-Selectivity
(CISS), had already been experimentally demonstrated by 
Ray et al.\cite{Ray99} as early as 1999,
but a real breakthrough took place with the two experimental studies
by Göhler et al.\cite{Gohler2011} and Xie et al.\cite{Xie11}, using
photoemission and AFM-based electrical transport probes, respectively.
Meanwhile it seems well confirmed that the CISS effect is a generic
feature of molecular systems displaying helical symmetry, although
a fully consistent theoretical description is still needed.\cite{Yeganeh09,Gutierrez12,Guo12,Medina12,Gutierrez13,Eremenko13,Rai13,0953-8984-26-1-015008,Guo14a,Guo14b,Medina15,Matityahu2015,Behnia16,Matityahu16,Caetano16,Diaz17,Wu17,Diaz17bis,doi:10.1021/acs.nanolett.9b01707,Michaeli19,doi:10.1021/acs.jpclett.9b02929} 
The vast majority of theoretical models proposed so far assume a close
connection between chirality and spin-orbit interactions in the molecular
systems, a result which seems to be supported by recent DFT-based calculations.\cite{doi:10.1021/acs.jpclett.8b02360,carmen}
However, part of the focus has been shifted recently to a more detailed
treatment of interface effects, which may play a non-trivial role,\cite{doi:10.1021/acs.nanolett.9b01707} as well as on correlation effects.\cite{doi:10.1021/acs.jpclett.9b02929}

Although most of the theoretical approaches are based on tight-binding
formulations, few of them started from a continuum formulation of
the problem.\cite{Yeganeh09,Medina15,Gutierrez13,Michaeli19}
When starting with a continuum model, it is common to simplify it
by restricting the electron's movement to a curved path and thus reducing
its spatial degrees of freedom to one, either by using a quantized
version of the classical Hamilton function for a particle moving on
a curve\cite{Medina15}, or by assuming an infinitely strong confinement
potential transversal to the helical path, leading to an exactly $1$-dimensional
effective Hamiltonian.\cite{Gutierrez13} In contrast, in Ref.~\cite{Michaeli19}
the exact eigenfunctions of an assumed transverse harmonic potential
were used in order to approximately map the 3D Hamiltonian onto an
effective 1D Hamiltonian (including an effective spin-orbit coupling). 
These transverse energy eigenstates (spinors) were 
labelled with an angular momentum index, thus keeping part
of the 3D nature of the model.
The resulting effective Hamiltonian, while providing a simple connection between geometry and 
spin-dependence, had some issues, such as the presence of terms proportional to an inverse power of the confinement length scale. This would imply arbitrarily large (spin-orbit) coupling  constants.

The quantization approach\cite{Medina15} has the advantage that no
confinement or projection is needed since the classical motion is
already restricted to the curve via holonomic constraints. However,
it is, from a formal mathematical point of view, not unambiguous\cite{Waldmann2007}
and the relation of the resulting description to the physics in the
ambient $3$-dimensional space is not very clear. 
Provided a well-defined limiting procedure exists, the approach based on a finite transversal 
confinement does not suffer from these problems. It is known that
the effective Hamiltonian does depend on extrinsic properties of the
curve (or more generally, the submanifold), i.\,e. its embedding
into the ambient space\cite{DaCosta1981} and thus cannot be obtained
by intrinsic quantization. This also means that the confined system
will retain some information about its surroundings.

Starting with the work of da Costa\cite{DaCosta1981}, adiabatic approximations of 
constrained quantum systems have been studied both in theoretical\cite{Maraner1995,Mitchell2001}
and mathematical physics\cite{Froese2001}. These results have been
generalized for potentials varying arbitrarily along the submanifold,\cite{Wachsmuth2014}
using the concept of adiabatic perturbation theory.\cite{Martinez2002,Sordoni2003,Panati2007,Teufel2003}
The works cited so far in this paragraph are all concerned with Schr\"odinger operators,
consisting only of a kinetic and a potential energy term. In order to model
spin-dependent processes in the absence of external magnetic fields, 
we need to include spin-orbit interactions,
which make decomposing the Hamiltonian into a longitudinal (or tangent)
and a transversal (or normal) contribution more complicated. 
An approximation scheme similar to da Costa's has been applied to systems 
including magnetic fields and spin-orbit coupling.\cite{Meijer2002,Zhang2007,Chang2013,Ortix2015}
However, an in-depth investigation concerning the applicability of this procedure in the presence of 
spin-orbit coupling is still missing.  

Taking as starting point and motivation the study of Ref.~\cite{Michaeli19},
we exploit the approach presented by Wachsmuth et. al.\cite{Wachsmuth2014}
to show that an adiabatic approximation is still possible for an electron
confined to a helix in the presence of spin-orbit coupling, and calculate
the effective Hamiltonian for different field configurations. Contrary
to Ref.~\cite{Michaeli19}, where the spin-orbit
interaction arose from the transversal confinement field, we consider 
a separate electric field as the source of spin-orbit coupling
which is different from the confinement. We lay our focus on fields
radially symmetric to the helix axis, but also give expressions for
helical fields and fields parallel to the helix axis (like those related
to an applied bias or the molecular dipole moment). Investigating
these different field configurations allows us to determine which types 
of spin-orbit coupling terms can occur in this kind of geometry without
using a fully general differential-geometric approach that might yield
less transparent results.

The adiabatic theory is used to show that physically well-motivated
approximations involving a separation ansatz are mathematically sound.
In this regard, this work can be seen as an application of an adapted
version of the adiabatic approximation to derive an explicit physical
model, which due to the inclusion of spin-orbit coupling lies beyond
the range of applications of this theory discussed so far in the literature. 
Considering a greater variety of spin-orbit coupling terms makes it easier for
us to determine whether the adiabatic approximation is applicable and
how the existing approach has to be altered to do so. In particular,
we find that the separation of the confinement and the spin-orbit
inducing field is necessary to fulfill the requirements of the adiabatic
method used here. Since our approach can be generalized to a broader
class of geometrical set-ups, we intend to clear the way forward to
the inclusion of spin-orbit coupling to the adiabatic description
of confined quantum systems.

We also provide a classification of the terms in our effective Hamiltonian 
by applying unitary transformations and rewriting it in terms of an 
effective gauge and magnetic field.
This allows for an easier comparison with other models and provides simple 
analytic expressions for the spin-related effective fields 
revealing their dependence on the model parameters. To round up our discussion, we also compute the spin polarization in the obtained effective model and show its dependence on various parameters. 

\section{The model}

To describe an electron in 3D space with helical confinement and a generic spin-orbit
coupling, we use the Pauli equation
\begin{equation}
\Heuc\Psi=-\frac{\hbar^{2}}{2m}\triangle\Psi+V+\Phi-i\frac{\hbar^{2}}{4m^{2}c^{2}}\boldsymbol{\sigma}\cdot\left(\nabla\Phi\times\nabla\Psi\right)\mmdot\label{eq:pauli}
\end{equation}
with a confinement potential $V$ and an additional scalar field $\Phi$. The potential $V$ rapidly  
increases in the directions normal to the helix (or some submanifold in general), thus restricting 
the particle motion to a small neighborhood of the helix, while the field $\Phi$ generates the 
spin-orbit coupling. The confinement can also be realized via homogeneous boundary conditions (which are equivalent to an infinitely deep potential well). 
$\boldsymbol{\sigma}=\left(\sigma_{x},\sigma_{y},\sigma_{z}\right)$
is the vector with the Pauli matrices as its entries and $\Psi$ is
a wave-function. The wave function $\Psi$ takes values in spin space
$\bbC^{2}$ (i.\,e. $\Psi\in L^{2}\left(\bbR^{3},\bbC^{2}\right)$
which is the space of square integrable $\bbC^{2}$-valued functions
on $3$-dimensional space).

The kinetic part of the Hamilton operator $\Hkine=-\frac{\hbar^{2}}{2m}\triangle+V+\Phi$
is of Schr\"odinger type and is diagonal in spin space. The remaining
part $\Hsoce$ is referred to as the spin-orbit-coupling (SOC) term. Notice that we are assuming that the confinement potential does not give rise to spin-orbit interaction terms; it just controls the extension of the electronic wave functions in the direction transversal to the helical path. The strength of the confinement is controlled by a confinement scale 
$\veps$ which the potential $V$ depends on (see Sec.~\ref{sec:confinement_pot}).

\subsection{Adapted local coordinates}\label{sec:adapted_coord}

\begin{figure}
\begin{centering}
\includegraphics[viewport=0bp 60bp 288bp 234bp,height=6cm]{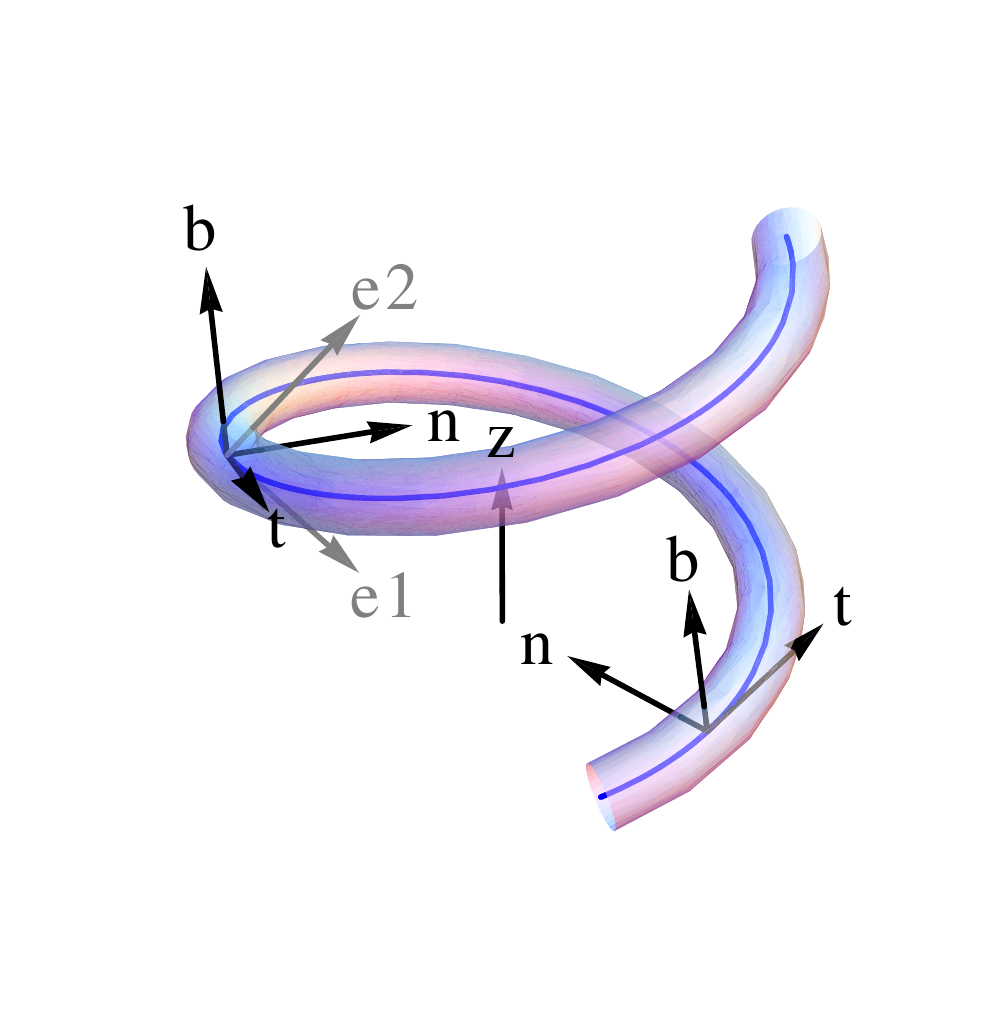}
\par\end{centering}
\caption{Helix with surrounding tube, Frenet frame 
$\left\{ \boldsymbol{t},\boldsymbol{n},\boldsymbol{b}\right\}$, 
rotated frame $\left\{ \boldsymbol{t},\boldsymbol{e}_1,\boldsymbol{e}_2\right\}$ 
and global $z$-axis.
\label{fig:helixtube}}
\end{figure}

When dealing with a confined system we need to choose suitable coordinates that cover 
a  sufficiently large neighborhood of the submanifold to which the particle motion is restricted. We 
consider a helix, which is 1-dimensional and can be described as an infinite Frenet curve $\boldsymbol{c}$ in 3D space with constant curvature and torsion (see Fig.~\ref{fig:helixtube}). 
Introducing local curvilinear coordinates, we obtain a map
\begin{equation}
\boldsymbol{r}\left(s,\rho,\varphi\right):=\boldsymbol{c}\left(s\right)+\rho\cos\varphi\boldsymbol{e}_{2}\left(s\right)+\rho\sin\varphi\boldsymbol{e}_{3}\left(s\right),\label{eq:parametrization}
\end{equation}
with 
\begin{equation}
\begin{aligned}\boldsymbol{e}_{2} & :=\cos\theta\boldsymbol{n}+\sin\theta\boldsymbol{b}\,\textrm{,}\\
\boldsymbol{e}_{3} & :=-\sin\theta\boldsymbol{n}+\cos\theta\boldsymbol{b},
\end{aligned}
\end{equation}
which represents a diffeomorphism from $\Omega_{\veps}=\bbR\times\left(0,\veps\right)\times\left(0,2\pi\right)$
to the tubular neighborhood $B_{\veps}$ of $\boldsymbol{c}$ for
all $\veps<1/\kappa$. The first coordinate of the new system is
simply the parameter of the curve $\boldsymbol{c}$, while $\rho$
and $\varphi$ are polar coordinates (with origin $\boldsymbol{c}\left(s\right)$)
in the plane normal to $\boldsymbol{t}\left(s\right)$. This is the most convenient choice 
if the confinement potential is spherically symmetric in the normal directions. 
Note that the function $\theta\left(s\right)$ fixes a certain rotation of the planar 
coordinate axes around $\boldsymbol{t}$ in each point $s$.

The metric tensor in these coordinates is given by:
\begin{equation}
g=\left(\begin{array}{ccc}
\left(1-\kappa\rho\cos\left(\varphi+\theta\right)\right)^{2}+\rho^{2}\left(\tau+\theta^{'}\right)^{2} & 0 & \rho^{2}\left(\tau+\theta^{'}\right)\\
0 & 1 & 0\\
\rho^{2}\left(\tau+\theta^{'}\right) & 0 & \rho^{2}
\end{array}\right)
\end{equation}
and
\begin{equation}
\det g=\rho^{2}\left(1-\kappa\rho\cos\left(\varphi+\theta\right)\right)^{2}\mmdot
\end{equation}

We choose
\begin{equation}
\theta\left(s\right)=-\integ[0][s]{\tau\left(\wtilde{s}\right)}{\wtilde{s}}\mmcom
\end{equation}
which implies that the local frame $\left\{ \frac{\der\boldsymbol{r}}{\der s},\frac{\der\boldsymbol{r}}{\der\rho},\frac{\der\boldsymbol{r}}{\der\varphi}\right\} $
is orthogonal on $\Omega_{\veps}$ and $g$ is a diagonal matrix. This choice is usually referred
to as a Tang frame.

The map $\boldsymbol{r}$ induces a transformation of the wave functions $\fun A{L^{2}\left(B_{\veps}\right)}{L^{2}\left(\Omega_{\veps}\right)}$
given by:
\begin{equation}
A\Psi:=\left(\det g\right)^{1/4}\Psi\circ\boldsymbol{r}\mmdot\label{eq:Atrans}
\end{equation}
The transformation $A$ is unitary and its inverse is:
\begin{equation}
\adj A\Psi=\left(\left(\det g\right)^{-1/4}\Psi\right)\circ\boldsymbol{r}^{-1}\mmdot
\end{equation}

The factor $\left(\det g\right)^{1/4}$ was introduced to absorb the
volume element coming from the curvilinear coordinates.

\subsection{The helix}

An explicit unit speed parametrization for the helix can be given as:
\begin{equation}
\boldsymbol{c}\left(s\right):=\left(R\cos\left(\frac{2\pi s}{\wtilde{R}}\right),R\sin\left(\frac{2\pi s}{\wtilde{R}}\right),\frac{bs}{\wtilde{R}}\right),
\end{equation}
with 
\begin{equation}
\wtilde{R}:=\pm\sqrt{\left(2\pi R\right)^{2}+b^{2}}\,\textrm{.}
\end{equation}
Using it, the corresponding Frenet frame reads:
\begin{equation}
\begin{aligned}\boldsymbol{t}\left(s\right) & =\left(-\frac{2\pi R}{\wtilde{R}}\sin\left(\frac{2\pi s}{\wtilde{R}}\right),\frac{2\pi R}{\wtilde{R}}\cos\left(\frac{2\pi s}{\wtilde{R}}\right),\frac{b}{\wtilde{R}}\right)\,\textrm{,}\\
\boldsymbol{n}\left(s\right) & =\left(-\cos\left(\frac{2\pi s}{\wtilde{R}}\right),-\sin\left(\frac{2\pi s}{\wtilde{R}}\right),0\right)\,\textrm{,}\\
\boldsymbol{b}\left(s\right) & =\left(\frac{b}{\wtilde{R}}\sin\left(\frac{2\pi s}{\wtilde{R}}\right),-\frac{b}{\wtilde{R}}\cos\left(\frac{2\pi s}{\wtilde{R}}\right),\frac{2\pi R}{\wtilde{R}}\right)\,\textrm{.}
\end{aligned}
\end{equation}
The helix has constant curvature and torsion which are related to its radius $R$ and pitch $b$ 
by the expressions:
\begin{equation}
\kappa=\frac{4\pi^{2}R}{\wtilde{R}^{2}}
\end{equation}
and
\begin{equation}
\tau=\pm\frac{b}{2\pi R}\kappa=\pm\frac{2\pi b}{\wtilde{R}^{2}}\mmdot
\end{equation}
It is worth noting that any curve with curvature and torsion both
nonzero and constant is a (left- or right-handed) helix, so this parametrization
describes all such curves. Therefore, the coordinates introduced in
\eqref{eq:parametrization} with $\kappa,\tau=\textrm{const}\neq0$
are called helical coordinates. Both $b$ and $\wtilde{R}$ change signs when 
the helicity is reversed (i.\,e. the helix is replaced by its mirrored counterpart).

Furthermore, the following abbreviations will be used later on:
\begin{equation}
\mathbf{u}_{\bot}\left(s,\rho,\varphi\right):=\rho\cos\left(\varphi+\theta\left(s\right)\right)\boldsymbol{n}\left(s\right)+\rho\sin\left(\varphi+\theta\left(s\right)\right)\boldsymbol{b}\left(s\right)
\end{equation}
which denotes all vectors orthogonal to the tangent vector $\boldsymbol{t}\left(s\right)$. As a result, the local parametrization becomes $\mathbf{x}=\boldsymbol{c}+\mathbf{u}_{\bot}$, while
\begin{equation}
\hat{n}=\boldsymbol{n}\cdot\mathbf{u}_{\bot}=\rho\cos\left(\varphi+\theta\left(s\right)\right)\,\textrm{, }\hat{b}=\boldsymbol{b}\cdot\mathbf{u}_{\bot}=\rho\sin\left(\varphi+\theta\left(s\right)\right)\label{eq:nhatbhat}
\end{equation}
are the projections of these normal vectors onto the Frenet frame.

\subsection{Confinement potential}\label{sec:confinement_pot}

The confinement potential $V$ in \eqref{eq:pauli} restricts the
particle motion to the proximity of the curve $c$. In curvilinear
coordinates we have: 
\begin{equation}
\left(AV\adj A\right)\left(s,\rho,\phi\right)=V\left(\boldsymbol{r}\left(s,\rho,\varphi\right)\right)=\veps^{-2}V_{0}\left(s,\veps^{-1}\rho,\vphi\right)\mmdot\label{eq:V0curv}
\end{equation}
Assuming that the shape of the potential does not vary
along the curve and that the potential is spherically symmetric, 
$V\circ\boldsymbol{r}$ only depends on $\rho$ and the scaling factor $\veps$ which 
controls the strength of the confinement. If $V$ is spherically symmetric, this 
can be imagined as the radius of a tube to which the electron motion is confined. 
As $\veps$ becomes small, the confinement increases while the tube radius shrinks.
We require that $V$ tends to $+\infty$ for $\rho\rightarrow\veps$, 
which implies that the wave functions obey 
homogeneous Dirichlet boundary conditions.\footnote{Provided the confinement potential $V\left(\boldsymbol{r}\left(s,\rho,\varphi\right)\right)=\veps^{-2}V_{0}\left(s,\veps^{-1}\rho,\vphi\right)$ is of order $\veps^{-4}$ (e.\,g. $V_{0}\left(\rho\right)=1/(2\veps^{4})\rho^{2}$), this is in fact not a limitation: Ref.~\cite[Prop. 8.1]{Froese2001} shows that under the initial condition $||\veps^{2}\triangle\Psi_{0}||^{2}\leq C$ the time evolution of $\Heuc$ approximates its counterpart with Dirichlet boundary conditions added.} 
This requirement is met by the potential well 
with walls of infinite height and radius $\veps$. 
This potential is zero within the tube
and thus realized by simply imposing the aforementioned boundary conditions 
on the wave functions.
The tube radius $\veps$ is also our confinement scale, so the limit
of strong confinement is equivalent to the tube having a radius close
to zero. Mathematically $\Heuc$ can be defined as the Friedrichs
extension\cite{Davies1995} of the operator in Eq.~\eqref{eq:pauli}
defined on $\ccinfty\left(B_{\veps}\right)$. The form domain of the
extension is $D\left(\sqrt{\Heuc}\right)=H_{0}^{1}\left(B_{\veps}\right)$.

So instead of a harmonic confinement as in Ref.~\cite{Michaeli19} we use
boundary conditions that are equivalent to an infinite-height two-dimensional
potential well in the normal plane of each point of the helix. This
potential has the advantage of being globally defined and also the
accessible space is covered by one single coordinate patch.

We also define rescaled operators by 
\begin{equation}
Q^{\veps}:=\adj[\veps]D\veps^{2}Q D_{\veps}\label{eq:epsscale}
\end{equation}
with $\left(D_{\veps}\Psi\right)\left(s,\rho,\vphi\right):=\veps^{-1}\Psi\left(s,\veps^{-1}\rho,\vphi\right)$.
The transformation \eqref{eq:epsscale} corresponds to switching to
microscopic time units $\wtilde t=\veps^{-2}t$ while at the same
time rescaling $V_{0}$ to become independent of $\veps$ which means
we have homogeneous boundary conditions on the cylinder $\Omega_{1}$ with
radius $1$ instead of $\Omega_{\veps}$.

\section{The effective Hamiltonian}

\subsection{\label{subsec:Kinetic-term}Transformation to helical coordinates}

As we have seen in Sec.~\ref{sec:adapted_coord}, the adapted local coordinates which are 
appropriate to describe our geometry are helical coordinates which give rise to the 
unitary transformation $A$ as defined in Eq.~\eqref{eq:Atrans}.
The relation between the Hamiltonian in euclidean and curvilinear (helical) coordinates is 
\begin{equation}
\Heuc\adj A\Psi=\adj A\Hcurv\Psi\label{eq:Heuccurveps}
\end{equation}
for $\Psi\in\ccinfty\left(\Omega_{\veps}\right)$ which extends to
the entire domain of $\Hcurv$ due to the uniqueness property of the
Friedrichs extension (see Sec.~III in the supplementary information).

We now rewrite the Hamilton operator \eqref{eq:pauli} using the curvilinear
coordinates \eqref{eq:parametrization}. We start with the kinetic
energy term in the Tang frame and obtain the following expression 
after applying the scaling transformation from Eq.~\eqref{eq:epsscale}:
\begin{equation}
\begin{aligned}\Hkinceps= & -\frac{\hbar^{2}}{2m}\left(\partdern{\rho}2+\frac{1}{\rho^{2}}\partdern{\varphi}2+\frac{1}{\left(1-\kappa\veps\hat{n}\right)^{2}}\veps^{2}\partdern s2+\frac{2\hat{b}\kappa\tau}{\left(1-\kappa\veps\hat{n}\right)^{3}}\veps^{3}\partder s\right.\\
 & \left.+\frac{\veps^{2}\kappa^{2}}{4\left(1-\kappa\veps\hat{n}\right)^{2}}+\frac{\kappa\tau^{2}\veps^{3}\left(5\veps\hat{b}^{2}\kappa-2\hat{n}\left(1-\kappa\veps\hat{n}\right)\right)}{4\left(1-\kappa\veps\hat{n}\right)^{4}}\right)\label{eq:hkinceps}
\end{aligned}
\end{equation}
$\Hkinceps$ can be written as the sum of a longitudinal part $H_{\textrm{kin},l}^{\veps}$
and a transversal part $H_{\textrm{kin},t}$ with the latter being
\begin{equation}
H_{\textrm{kin},t}=-\frac{\hbar^{2}}{2m}\left(\partdern{\rho}2+\frac{1}{\rho^{2}}\partdern{\varphi}2+\frac{1}{4\rho^{2}}\right),\label{eq:kin0tang}
\end{equation}
which is independent of $\veps$. The longitudinal part of $\Hkinceps$ now only depends
on partial derivatives with respect to the arc length $s$. Outside of $H_{\textrm{kin},t}$, 
partial derivatives with respect to the transverse coordinates $\rho$ and $\vphi$ still 
appear in the spin-orbit coupling term $\Hsocc$.

We assumed homogeneous boundary conditions for the tube $B_{\veps}$,
which is transformed and rescaled to the cylinder $\Omega_{1}=\bbR\times B_{2}\left(0,1\right)$
(in cylinder coordinates). Therefore $H_{\textrm{kin},t}$ is proportional
to the $2$-dimensional Dirichlet-Laplace operator $-\triangle$ on
the $2$d-ball $B_{2}\left(0,1\right)$ with radius $1$. Later we
will also explicitly write down $\Hsocc$, the spin-orbit coupling
term in helical coordinates.

For the following two sections we assume initial conditions $\Psi_{0}$ with $\Norm{\Psi_{0}}=1$ and $\Norm{\veps^{2}\triangle\Psi_{0}}^{2}\leq C$ which is 
sufficient to show adiabatic approximation. In Section \ref{subsec:transversal_states} 
we will restrict ourselves to initial conditions with 
$\Norm{\veps^{2}\triangle\Psi_{0}}^{2}\leq C\veps^2$ in order to obtain 
a simplified formula for the approximate Hamiltonian allowing for its direct 
computation using $\Hcurv$.\footnote{Using 
$||\veps^{2}\triangle\Psi_{0}||^{2}\leq C\veps^2$ instead of 
$||\veps^{2}\triangle\Psi_{0}||^{2}\leq C$ simplifies the adiabatic construction if 
no spin-orbit coupling terms are present. Including SOC however negates this advantage, 
so we might as well start with the more general condition and strengthen it at a later 
point if necessary.}
Furthermore, the Taylor expansion of $\Hcurveps$ up to second order in $\veps$
shall be denoted by 
\begin{equation}
H^{\veps}=\Hkineps+\Hsoceps=H_{0}+\veps H_{1}+\veps^{2}H_{2}\mmdot
\end{equation}
We observe that for states $\Psi\in\chi_{\left(-\infty,E\right]}\left(H^{\veps}\right)L^{2}\left(\Omega_{\delta}\right)$,
which is the subspace with energy cut-off at $E$, the Taylor expansion
$H^{\veps}$ of $\Hcurveps$ differs from $\Hcurveps$ itself only
by an error of order $3$: 
\begin{equation}
\Norm{\left(\Hcurveps-H^{\veps}\right)\Psi}_{D\left(H^{\veps}\right)}\leq\wtilde C\veps^{3}\mmdot\label{eq:taylorapprox}
\end{equation}
That this approximation is also valid for the time evolution generated
by $\Hcurveps$ and $H^{\veps}$ can be shown by transferring the
arguments in the proof of Ref.~\cite[Corollary 1]{Wachsmuth2014}
to our situation.\footnote{Replace $H_{\mathcal{A}}^{\veps}$ with $\Hcurveps$ and $H_{\textrm{eff}}^{\veps}$
with $\tilde{H}_\textrm{eff}$ and use Eq.~\eqref{eq:adiabappr}.} 
Therefore, we can use the second order Taylor expansion $H^{\veps}$
of $\Hcurveps$ instead of the full expression.

\subsection{Transversal solutions}

We shall now determine the eigenfunctions of the transverse Hamiltonian 
$H_{\textrm{kin},t}$, together with the boundary condition
which requires the solutions to vanish on the boundary of the unit
circle (after the rescaling \eqref{eq:epsscale}). Those are required to calculate
the effective (approximate) Hamiltonian below. Being a Dirichlet-Laplacian,
$H_{\textrm{kin},t}$ has a purely discrete spectrum,\cite{Davies1995}
implying that all eigenenergies are isolated points. Since the transverse
part does not depend on spin, we omit spin-degrees of freedom in this
section entirely, thus dealing with complex-valued wave functions.

Using the ansatz 
\begin{equation}
\psi_{N,l}\left(\rho,\vphi\right)=\sqrt{k_{N,\left|l\right|}\rho}J_{\left|l\right|}\left(k_{N,\left|l\right|}\rho\right)e^{il\vphi}
\end{equation}
with $k_{N,\left|l\right|}=\frac{\sqrt{2mE_{N,\left|l\right|}}}{\hbar}$,
we obtain for $x=k_{N,\left|l\right|}\rho$: 
\begin{equation}
x^{2}J_{\left|l\right|}^{''}\left(x\right)+xJ_{\left|l\right|}^{'}\left(x\right)+\left(x^{2}-l^{2}\right)J_{\left|l\right|}\left(x\right)=0\mmcom
\end{equation}
i.\,e., the $J_{\left|l\right|}$ are solution's of Bessel's differential
equation. Since $\psi$ has to be square integrable, the solutions
$J_{\left|l\right|}$ are Bessel functions of the first kind. The
energy $E_{N,\left|l\right|}$ is determined by the boundary condition
$J_{\left|l\right|}\left(k_{N,\left|l\right|}\right)=0$ to 
\begin{equation}
E_{N,\left|l\right|}=\frac{\hbar^{2}}{2m}\left(j_{\left|l\right|,N}\right)^{2}
\end{equation}
where $j_{\left|l\right|,N}$ is the $N$-th zero of the Bessel function
$J_{\left|l\right|}$. $l$ is an integer due to the periodicity of
$\psi_{N,l}$ while $J_{\left|l\right|}$ only depends on the absolute
value of $l$. The quantum number $l$ labels the angular momentum
part of the full wave function. The lowest enery with $l\neq0$ is
$E_{1,1}$ with eigenstates $\psi_{1,1}$ and $\psi_{1,-1}$.
The projection to the subspace of the $L^{2}\left(\Omega_{1}\right)$-functions
with the transverse part lying in the eigenspace of $E_{1,1}$ is
defined as: 
\begin{equation}
P_{0}:=\ident\otimes\ket{\psi_{1,1}}\bra{\psi_{1,1}}+\ident\otimes\ket{\psi_{1,-1}}\bra{\psi_{1,-1}}\mmdot
\end{equation}
We also define a map $\fun{U_{0}}{L^{2}\left(\Omega_{\delta}\right)}{L^{2}\left(\bbR,\bbC^{2}\right)}$
whose restriction to $P_{0}L^{2}\left(\Omega_{\delta}\right)$ is
unitary, by: 
\begin{equation}
\left(U_{0}\Psi\right)_{l}\left(s\right):=\integ[0][\delta]{{\integ[0][2\pi]{\psi_{1,l}^{\ast}\left(\rho,\vphi\right)\Psi\left(s,\rho,\vphi\right)}{\vphi}}}{\rho},\label{eq:U0def}
\end{equation}
with the adjoint 
\begin{equation}
\left(\adj[0]Uf\right)\left(s,\rho,\vphi\right)=\sum_{l\in\left\{ -1,1\right\} }\psi_{1,l}\left(\rho,\vphi\right)f_{l}\left(s\right)\mmdot
\end{equation}
This operator obeys the relations $U_{0}\adj[0]U=\ident$ and $\adj[0]UU_{0}=P_{0}$.
Note that the subspace which $U_{0}$ maps into is $L^{2}\left(\bbR,\bbC^{2}\right)$,
the space of square-integrable functions on the real line with values
in $\bbC^{2}$ which does not describe spin (which is omitted here)
but angular momentum orientation. So if we include spin again, we
are dealing with $\bbC^{2}\times\bbC^{2}$-valued functions, which
have both a spin and an angular momentum index.

\subsection{\label{subsec:Adiabatic_approx}Adiabatic approximation}

We now aim at finding an approximation for the Hamiltonian $\Hcurveps$
in the limit of a strong confinement potential. The potential $V_{0}$
in Eq.~\eqref{eq:V0curv} scales the directions normal to the helix
by $\veps^{-1}$, so that the confinement becomes strong for small
$\veps$. This means the potential well we use for confinement has radius
$\veps$ and becomes arbitrarily narrow if $\veps$ goes to zero.
This limit has structural similarities to the Born-Oppenheimer approximation
with the transversal scaling factor $\veps$ taking the role of the
electron to nucleus mass ratio. Therefore, ideas from adiabatic perturbation
theory can be taken over to the case of strong confining forces.

We are looking for an effective Hamiltonian which is defined on the
reduced space\\
$U_{0}L^{2}\left(\Omega_{1},\bbC^{2}\right)=L^{2}\left(\bbR,\bbC^{2}\times\bbC^{2}\right)$.
Based on our previous considerations, an educated guess would be 
\begin{equation}
H_{\textrm{eff}}^{\left(0\right)}=U_{0}H^{\veps}\adj[0]U\mmdot\label{eq:Heffdef}
\end{equation}
Indeed, for vanishing curvature and torsion and without spin-orbit
coupling the $H_{\bbR^{3}}$ can be written as a sum of a purely transversal
and a purely longitudinal contribution, allowing for a simple separation
ansatz. In other words $\left[H_{\bbR^{3}},P_{0}\right]=0$ would
hold and
\begin{equation}
\left(\adj[0]UH_{\textrm{eff}}^{\left(0\right)}U_{0}-H_{\bbR^{3}}\right)P_{0}=0\label{eq:HeffnoSOC}
\end{equation}
would follow. However, both the SOC term and the curved geometry spoil
this simple behaviour and Eq.~\eqref{eq:HeffnoSOC} is no longer true
in general. Therefore, a careful estimate for the error introduced
by replacing $\Hcurveps$ with some effective Hamiltonian is needed.
In the following we sketch a way to get such an estimate using adiabatic
perturbation theory.

Following the approach in Ref.~\cite{Wachsmuth2014} one starts by 
constructing a closed subspace of $L^{2}\left(\Omega_{1}\right)$ which 
is invariant up to some small error under the dynamics generated
by $H^{\veps}$. We call the orthogonal projector associated with this subspace 
$P_{\veps}$ and require the properties:\footnote{Here and in the following, the domain $D\left(H\right)$ of an operator
$H$ is always assumed to be equipped with the graph norm. If $H$ is closed, 
$D\left(H\right)$ is a Hilbert space whose scalar product induces the graph norm and 
$H\in\mathcal{L}\left(D\left(H\right),\mathcal{H}\right)$ holds.}
\begin{enumerate}
\item $P_{\veps}$ is bounded on $D\left(\left(H^{\veps}\right)^{m}\right)$
and $P_{\veps}-P_{0}=\bigO\left(\veps\right)$ as a bounded operator
on $D\left(\left(H^{\veps}\right)^{m}\right)$, 
\item $\left[H^{\veps},P_{\veps}\right]=\bigO\left(\veps\right)$ as a bounded
operator from $D\left(\left(H^{\veps}\right)^{m+1}\right)$ to $D\left(\left(H^{\veps}\right)^{m}\right)$
and 
\item $\left[H^{\veps},P_{\veps}\right]\chi_{\left(-\infty,E\right]}\left(H^{\veps}\right)=\bigO\left(\veps^{3}\right)$
as a bounded operator from $L^{2}\left(\Omega_{1}\right)$ to $D\left(\left(H^{\veps}\right)^{m}\right)$.
\end{enumerate}
The construction of $P_{\veps}$ is sketched in Sec.~IV of the supplementary information.

By setting
\begin{equation}
U^{\veps}:=\left(P_{0}P_{\veps}+P_{0}^{\bot}P_{\veps}^{\bot}\right)\left(\ident-\left(P_{\veps}-P_{0}\right)^{2}\right)^{-1/2}
\end{equation}
we obtain a unitary map
$\fun{U^{\veps}}{L^{2}\left(\Omega_{1}\right)}{L^{2}\left(\Omega_{1}\right)}$
whose restriction to $P_{\veps}L^{2}\left(\Omega_{1}\right)$ is also
unitary as a map from $P_{\veps}L^{2}\left(\Omega_{1}\right)$ to
$P_{0}L^{2}\left(\Omega_{1}\right)$. $U^{\veps}$ is bounded as an
operator on $D\left(\left(H^{\veps}\right)^{m}\right)$ and admits
an expansion 
\begin{equation}
U^{\veps}=\ident+\veps U_{1}^{\veps}+\veps^{2}U_{2}^{\veps}\label{eq:Uepsseries}
\end{equation}
with the $U_{i}^{\veps}$ being of order $\veps^{0}$ (as operators
on $D\left(\left(H^{\veps}\right)^{m}\right)$) and $P_{0}U_{1}^{\veps}P_{0}=0$
and $U_{2}^{\veps}P_{0}=P_{0}U_{2}^{\veps}P_{0}=P_{0}U_{2}^{\veps}$.

Using $U_{0}$ and $U^{\veps}$ we define  an operator $U:=U_{0}U^{\veps}$
with the properties: 
\begin{enumerate}
\item $U$ is a unitary map from $P_{\veps}\mathcal{H}$
to $L^{2}\left(\bbR^{d},\bbC^{2}\times\bbC^{2}\right)$, 
\item $U\adj U=\ident_{L^{2}\left(\bbR^{d}\right)}$ and $\adj UU=P_{\veps}$.
\end{enumerate}
Setting $\Hefftilde=UH^{\veps}\adj U$ we can show that
\begin{equation}
\Norm{\Bigl(e^{-itH^{\veps}}-{{\adj U}}e^{-it\Hefftilde}U\Bigr)P_{\veps}\chi_{\left(-\infty,E\right]}\left(H^{\veps}\right)}\leq C_{1}\veps^{3}\left(1+\left|t\right|\right)\label{eq:adiabappr}
\end{equation}
i.\,e. on the subspace $P_{\veps}L^{2}\left(\Omega_{1}\right)$
with cut-off energy $E$, $H^{\veps}$ and $\Heff$ yield approximatly
the same dynamics up to an error of order $\veps^{3}$ (and up to
an error of $\veps$ on the macroscopic time scale or times of order
$\veps^{-2}t$ respectively). 
This error estimate which follows from property 3. of $P_\veps$ by Duhamel's principle 
as in Ref.~\cite[Eq.~(31)]{Wachsmuth2014} is our main technical result. 
It shows that the deviation between the time evolution of the effective and the full 
Hamiltonian is controlled by the adiabatic scale. This approximation is sometimes called a 
superadiabatic\cite{Wachsmuth2014,Joye1993} since the estimate depends on a power of $\veps$ 
greater than one. Likewise 
$\exp\left(-it\wtilde{H}^\veps\right)$ and $P_\veps L^2\left(\Omega_1\right)$ are called 
superadiabatic evolution and subspace respectively. 
Estimates like \eqref{eq:adiabappr} can be 
obtained without this super-adiabatic construction but with a power of $\veps$ below 3 
which is insufficient for approximation on macroscopic 
time scales (see Ref.~\cite[Sec.~1.2]{Wachsmuth2014}).

While Eq.~\eqref{eq:adiabappr} gives us an error estimate for the
adiabatic approximation, the effective Hamiltonian still differs from
the guess we made in \eqref{eq:Heffdef}, which is more straightforward
to calculate. However since $U^{\veps}$ is a second order polynomial
in $\veps$ by \eqref{eq:Uepsseries}, an additional argument similar
to Ref.~\cite[p.~52ff.]{Wachsmuth2014} yields that we can neglect
all higher order terms in \eqref{eq:adiabappr} and $\Hefftilde=U_{0}U^{\veps}H^{\veps}\adj{\left(U^{\veps}\right)}\adj[0]U$
except for the expression 
\begin{equation}
H_{\textrm{corr}}=\veps^{2}U_{0}H_{1}B_{e}H_{1}\adj[0]U\label{eq:problemterm}
\end{equation}
with $B_{e}=\left(E_{0}-H_{\textrm{kin},t}^{\bot}\right)^{-1}P_{0}^{\bot}$
and $H_{\textrm{kin},t}^{\bot}$ being the restriction of $H_{\textrm{kin},t}$
to $P_{0}^{\bot}L^{2}\left(\Omega_1\right)$:

We can show 
\begin{equation}
\Norm{\left(\Hefftilde-\Heff\right)\chi_{\left(-\infty,E\right]}\left(\Hefftilde\right)}\leq C_{2}\veps^{3}
\end{equation}
with 
\begin{equation}
\Heff=H_{\textrm{eff}}^{\left(0\right)}+H_{\textrm{corr}}\mmdot\label{eq:Heffcorr}
\end{equation}
As in Ref.~\cite[Corollary 2, see also p. 44]{Wachsmuth2014} we can
additionally replace $U$ by $U_{0}$ in \eqref{eq:adiabappr} while
only aqcuiring an additional time-independent error of order $\veps$
to obtain\footnote{Most of the proof carries over to our situation by replacing $H_{\textrm{eff}}^{\left(2\right)}$
with $\Heff$, $H_{\textrm{eff}}^{\veps}$ with $\tilde{H}_\textrm{eff}$ and
$U^{\veps}$ with $U$. $U-U_{0}=\bigO\left(\veps\right)$ in $\mathcal{L}\left(L^{2}\left(\Omega_{1}\right),L^{2}\left(\bbR\right)\right)$
follows directly from \eqref{eq:Uepsseries}.}
\begin{equation}
\Norm{\Bigl(e^{-itH^{\veps}}-{{\adj[0]U}}e^{-it\Heff}U_{0}\Bigr){{\adj[0]U}}\chi_{\left(-\infty,E\right]}\left(\Heff\right)U_{0}}\leq C_{3}\veps\left(1+\veps^{2}\left|t\right|\right)
\end{equation}
with $\Heff=H_{\textrm{eff}}^{\left(0\right)}+H_{\textrm{corr}}$
as in \eqref{eq:Heffcorr}. One might ask why those higher order terms
were required in the first place. The answer is that we needed $\left[H^{\veps},P_{\veps}\right]\chi_{\left(-\infty,E\right]}\left(H^{\veps}\right)=\bigO\left(\veps^{3}\right)$
to get the third order error estimate in \eqref{eq:adiabappr}. Using
$P_{0}$ instead of $P_{\veps}$ together with $H_{\textrm{eff}}^{\left(0\right)}=U_{0}H^{\veps}\adj[0]U$
would only give an error estimate of order $\veps$ because of $\left[H^{\veps},P_{0}\right]=\bigO\left(\veps\right)$
due to the spin-orbit coupling terms and the higher order kinetic terms coming from the 
helical geometry.

We will now explicitly write down the kinetic part of $H_{\textrm{eff}}^{\left(0\right)}$. The
scalar product $\braket{\psi_{1,l}|H^{\veps}\psi_{1,l^{'}}}$ involves 
integrations over $\varphi$ and $\rho$. Because $\psi_{1,l^{'}}$
is an eigenvector of $-i\partder{\varphi}$, the latter is always
replaced by $l^{'}$. From our calculation of $\Hcurv$, we now that the prefactors of the derivatives are rational functions of $\hat{b}$ and $\hat{n}$ (see Eq.~
\eqref{eq:hkinceps}). So $H_{\textrm{eff}}^{\left(0\right)}$ consists of expressions of the form
\begin{equation}
\integ[0][\delta]{{\integ[0][2\pi]{\psi^*_{1,l}\left(\rho,\varphi\right)\rho^{k+n+m}\cos^{n}\varphi\sin^{m}\varphi{{\partdern{\rho}r}}\psi_{1l^{'}}\left(\rho,\varphi\right)}{\varphi}}}{\rho}\label{eq:genint}
\end{equation}
multiplied with functions of and partial derivatives with respect
to $s$. Due to \eqref{eq:nhatbhat} and the $\varphi$-dependence
of $\psi_{1,l}$, \eqref{eq:genint} vanishes for uneven values $n+m$
because of $l\in\left\{ -1,1\right\} $ and the orthogonality of the
eigenfunctions. In the following, we therefore neglect terms with
an uneven number of Sine and Cosine because these vanish upon projection anyway and 
therefore do not contribute to $H_{\textrm{eff}}^{\left(0\right)}$.
So by taking the Taylor expansion of $\Hkinceps$ and omitting uneven powers of 
the trigonometric functions, we obtain for the kinetic term: 
\begin{equation}
\Hkineps=-\frac{\hbar^{2}\epsilon^{2}}{2m}\left(\partdern s2+\frac{\kappa^{2}}{4}\left(1+3\rho^{2}\epsilon^{2}\left(\ell_{-}+\ell_{+}+2\right)\partdern s2\right)\right)+H_{\textrm{kin},t}
\end{equation}
where the definitions 
\begin{align}
\ell_{+} & :=\exp\left(i2\varphi\right)\textrm{ and }\ell_{-}:=\exp\left(-i2\varphi\right)\label{eq:lupdowndef}
\end{align}
were used.

\subsection{Spin-orbit coupling}

We assume that the spin-orbit coupling is induced by a scalar potential
$\wtilde{\Phi}$ leading to a radial field around the (global) $x^{3}$-axis,
i.\,e. 
\begin{equation}
\wtilde{\Phi}\left(\mathbf{x}\right)=\Phi\left(\sqrt{\left(x^{1}\right)^{2}+\left(x^{2}\right)^{2}}\right)\,\textrm{.}\label{eq:zradialfield}
\end{equation}
As for $\Hkineps$, the spin-orbit coupling term $\Hsoceps$ is obtained by calculating the Taylor expansion of the full 
expression for the SOC up to second order in $\veps$. We again omit uneven powers of Sine and 
Cosine because they vanish upon projection anyway. Using again the definition \eqref{eq:lupdowndef} this results in:
\begin{equation}
\Hsoceps=\frac{\hbar^{2}\veps^{2}}{16m^{2}c^{2}}\left(\Phi^{'}\left(R\right)h_{01}+\Phi^{''}\left(R\right)h_{02}\right)\label{eq:hsocfull}
\end{equation}
with
\begin{equation}
\begin{aligned} & h_{01}=4\boldsymbol{\sigma}\cdot\boldsymbol{b}i\partder s\\
 & +\left(\frac{b^{2}\boldsymbol{\sigma}\cdot\boldsymbol{t}\left(\ell_{-}-\ell_{+}\right)}{R\wtilde R^{2}}+\boldsymbol{\sigma}\cdot\boldsymbol{b}\tau\left(\ell_{-}-\ell_{+}\right)-i\boldsymbol{\sigma}\cdot\boldsymbol{n}\tau\left(\ell_{-}+\ell_{+}-2\right)\right)\rho\partder{\rho}\\
 & +\left(\frac{b^{2}\boldsymbol{\sigma}\cdot\boldsymbol{t}\left(\ell_{-}+\ell_{+}-2\right)}{R\wtilde R^{2}}+\boldsymbol{\sigma}\cdot\boldsymbol{b}\tau\left(\ell_{-}+\ell_{+}-2\right)+i\boldsymbol{\sigma}\cdot\boldsymbol{n}\tau\left(\ell_{+}-\ell_{-}\right)\right)\left(-i\partder{\vphi}\right)\\
 & +\frac{b^{2}\boldsymbol{\sigma}\cdot\boldsymbol{t}\left(\ell_{+}-\ell_{-}\right)}{2R\wtilde R^{2}}+\frac{1}{2}\boldsymbol{\sigma}\cdot\boldsymbol{b}\tau\left(\ell_{+}-\ell_{-}\right)+\frac{1}{2}i\boldsymbol{\sigma}\cdot\boldsymbol{n}\tau\left(\ell_{-}+\ell_{+}-2\right)
\end{aligned}
\end{equation}
and 
\begin{equation}
h_{02}=\boldsymbol{\sigma}\cdot\boldsymbol{t}\left(\ell_{+}-\ell_{-}\right)\rho\partder{\rho}-\boldsymbol{\sigma}\cdot\boldsymbol{t}\left(\ell_{-}+\ell_{+}+2\right)\left(-i\partder{\vphi}\right)+\frac{1}{2}\boldsymbol{\sigma}\cdot\boldsymbol{t}\left(\ell_{-}-\ell_{+}\right)\mmdot
\end{equation}
Since the full SOC Hamiltonian is self-adjoint, this series expansion
is self-adjoint as well, which can also be checked by direct computation.

It is instructive to consider potentials with different symmetries
as well. If we choose 
\begin{equation}
\wtilde{\Phi}\left(\mathbf{x}\right)=-E_{z}x^{3}
\end{equation}
i.\,e. a constant external field parallel to the global $x^{3}$-axis which could be the result 
of a molecular dipole moment or an applied bias, we obtain 
\begin{equation}
\begin{aligned}\Hsoceps & =\frac{i\hbar^{2}\veps^{2}}{2m^{2}c^{2}}E_{z}\Biggl(\frac{\pi R}{\wtilde R}\left(\sigma_{x}\cos\left(\frac{2\pi s}{\wtilde R}\right)+\sigma_{y}\sin\left(\frac{2\pi s}{\wtilde R}\right)\right)\partder s\\
 & \hphantom{=\frac{i\hbar^{2}\veps^{2}}{2m^{2}c^{2}}E_{z}\Biggl(}+\frac{\pi^{2}R}{\wtilde R^{2}}\left(-\sigma_{x}\sin\left(\frac{2\pi s}{\wtilde R}\right)+\sigma_{y}\cos\left(\frac{2\pi s}{\wtilde R}\right)\right)\Biggr)
\end{aligned}
\label{eq:hsoczdir}
\end{equation}
which reproduces the result from Ref.~\cite[Eq.~(2)]{Medina15}.

Another example is a field which is radially symmetric to the helix,
i.\,e. it only depends on the normal distance parameter $\rho$:
\begin{equation}
\wtilde{\Phi}\left(\boldsymbol{x}\right)=\Phi\left(\rho\right)\mmdot
\end{equation}
In this case the spin-orbit coupling reads 
\begin{equation}
\Hsoceps=\frac{\hbar^{2}}{4m^{2}c^{2}}\left(-\Phi^{''}\left(0\right)\epsilon^{2}l\boldsymbol{\sigma}\cdot\boldsymbol{t}+\Phi^{'}\left(0\right)\left(-\frac{l\epsilon\boldsymbol{\sigma}\cdot\boldsymbol{t}}{\rho}-\frac{1}{2}i\epsilon^{3}\kappa\rho\boldsymbol{\sigma}\cdot\boldsymbol{b}\partder s\right)\right)\mmdot\label{eq:Hsockarenplus}
\end{equation}
This expression differs from \eqref{eq:hsocfull} and \eqref{eq:hsoczdir}
in that terms occur at different orders of $\veps$. In case of $\Phi^{'}\left(0\right)=0$
the SOC term is of the same form ($\propto l\boldsymbol{\sigma}\cdot\boldsymbol{t}$)
as in Ref.~\cite{Michaeli19} where the helical confinement potential acted as the source of 
the SOC.

$s$-dependent versions of all these fields can be considered as well; 
this leads to additional terms proportional to $\sigma_{x}$ as well as $s$-dependent 
prefactors. Expressions for $s$-dependent fields are given in Sec.~V of the supplementary information.

Had we used the confinement potential as the source for SOC as in
Ref.~\cite{Michaeli19}, we would have run into an issue here: in this
case $\Phi\left(\rho\right)=\frac{\hbar^{2}}{2m}\frac{\rho^{2}}{\veps^{4}}$
would hold and therefore
\begin{equation}
-\frac{\hbar^{4}}{4m^{3}c^{2}\veps^{2}}l\boldsymbol{\sigma}\cdot\boldsymbol{t}
\end{equation}
would be the leading order term in $\Hsoceps$. This is a contribution
of order $\veps^{-2}$ which does not fit into the adiabatic approach
presented here. Since the term becomes arbitrarily large for small $\veps$
it will be difficult to include it in any systematic approximation procedure dealing
with the limit $\veps\rightarrow0$.

\subsection{\label{subsec:transversal_states}Projection with transversal states}

We write down the particular cases of \eqref{eq:genint} that
we need to calculate the effective Hamiltonian $H_{\textrm{eff}}^{\left(0\right)}=H_{\textrm{kin},\textrm{e}}+H_{\textrm{SOC},\textrm{e}}$:
\begin{equation}
\begin{aligned}\integ[0][2\pi]{{\integ[0][1]{\psi^*_{1,l}\left(\rho,\varphi\right)\psi_{1,l^{'}}\left(\rho,\varphi\right)}{\rho}}}{\vphi} & =\delta_{ll^{'}}\textrm{,}\\
\integ[0][2\pi]{{\integ[0][1]{\psi^*_{1,l}\left(\rho,\varphi\right)\rho{{\partder{\rho}}}\psi_{1,l^{'}}\left(\rho,\varphi\right)}{\rho}}}{\vphi} & =-\frac{1}{2}\delta_{ll^{'}}\mmcom\\
\integ[0][2\pi]{{\integ[0][1]{\psi^*_{1,l}\left(\rho,\varphi\right)\rho^{2}\psi_{1,l^{'}}\left(\rho,\varphi\right)}{\rho}}}{\vphi} & =\frac{1}{3}
\end{aligned}
\end{equation}
for $l\in\left\{ -1,1\right\}$.

Using these integrals we obtain
\begin{equation}
\begin{aligned}\left(H_{\textrm{kin},\textrm{e}}\right)_{ll^{'}} & =E_{1,1}-\frac{\hbar^{2}\veps^{2}}{2m}\Biggl(\left(\partdern s2-\frac{\kappa^{2}}{4}\left(2\epsilon^{2}\partdern s2+1\right)\right)\delta_{ll^{'}}\\
 & -\frac{\epsilon^{2}\kappa^{2}}{4}\partdern s2\left(\delta_{ll^{'}-2}+\delta_{ll^{'}+2}\right)\Biggr)
\end{aligned}
\label{eq:Hkine}
\end{equation}
and 
\begin{equation}
\begin{aligned}\left(H_{\textrm{SOC},\textrm{e}}\right)_{ll^{'}} & =\frac{\hbar^{2}\veps^{2}}{8m^{2}c^{2}}\Biggl(\Phi^{'}\left(R\right)\left(-\frac{b^{2}l\boldsymbol{\sigma}\cdot\boldsymbol{t}}{R\wtilde R^{2}}+\boldsymbol{\sigma}\cdot\boldsymbol{b}(-l\tau+2i\partder s)-i\boldsymbol{\sigma}\cdot\boldsymbol{n}\tau\right)\\
 & \hphantom{=\frac{\hbar^{2}\veps^{2}}{8m^{2}c^{2}}\Biggl(}+\Phi^{''}\left(R\right)\left(-l\boldsymbol{\sigma}\cdot\boldsymbol{t}\right)\Biggr)\delta_{ll^{'}}\mmdot
\end{aligned}
\label{eq:Hsoce}
\end{equation}
For $\Phi^{'}\left(R\right)=0$ (e.\,g. if the radial field as a minimum at the 
radius of the helix) we again obtain an SOC term proportional to
$-l\boldsymbol{\sigma}\cdot\boldsymbol{t}$ as in Ref.~\cite{Michaeli19}

At this point we note that unlike $H_{\textrm{SOC},\textrm{e}}$,
$H_{\textrm{kin},\textrm{e}}$ is not proportional to $\delta_{ll^{'}}$.
However we can get rid of the off-diagonal term by using a more restricted initial condition. 
So instead of an initial state $\Psi_{0}$ with $\Norm{\veps^{2}\partdern[\Psi_{0}]s2}\leq C$ 
we shall use a state with $\Norm{\veps^{2}\partdern[\Psi_{0}]s2}\leq C\veps^2$ from now on,\footnote{The initial condition introduced in Sec.~\ref{subsec:Kinetic-term} allowed for kinetic energies of the same order as the confinement potential, see Eq.~\eqref{eq:V0curv}. With the revised initial condition, the kinetic energy has to be small compared to the confinement thus narrowing the energy range where the approximation is justified.} 
i.\,e. the kinetic energy of $\Psi_0$ is of order $\veps^2$ instead of $\veps^0$. 
The off-diagonal term will then be of order $\veps^{4}$ and can therefore be neglected 
(see Eq.~\eqref{eq:taylorapprox}). 

We still have to address how to deal with the second term $H_{\textrm{corr}}$
in \eqref{eq:Heffcorr}. It depends on the resolvent map $\left(E_{0}-H_{\textrm{kin},t}^{\bot}\right)^{-1}$
and the first order part $H_{1}$ of the Hamiltonian which reads:
\begin{equation}
\begin{aligned} & H_{1}=-\frac{\hbar^{2}}{m}\kappa\rho\cos\phi\epsilon^{2}\partdern s2\\
 & +\frac{\hbar^{2}}{4m^{2}c^{2}}\Phi^{'}\left(R\right)i\left(\boldsymbol{\sigma}\cdot\boldsymbol{t}\left(\frac{\sin\phi}{2\rho}-\sin\phi\partder{\rho}-\frac{\cos\phi}{\rho}\partder{\vphi}\right)+\boldsymbol{\sigma}\cdot\boldsymbol{b}\epsilon\partder s\right)\mmdot
\end{aligned}
\label{eq:Hfirstorder}
\end{equation}
 With our updated initial condition we can neglect all the terms coming from $H_{\textrm{corr}}$ 
 involving partial derivatives. The only remaining term is the one quadratic in the spin-orbit
coupling constant $\frac{\hbar^{2}}{4m^{2}c^{2}}\Phi^{'}\left(R\right)$,
which is a negligibly small correction to the first order spin-orbit
coupling. The remainder of $\Heff$ is now of the form \eqref{eq:Heffdef}, i.\,e. 
the second order Taylor expansion of the Hamiltonian in helical coordinates 
projected with the transversal states.
Finally, the potential $\Phi$ which was included for consistency in
Eq.~\eqref{eq:pauli} is up to second order in $\veps$ just the constant
$\Phi\left(R\right)$, which leads to a trivial energy shift upon
projection.
The remaining kinetic and SOC terms are of second
order in $\veps$ and we can simply divide them by $\veps^{2}$ to
switch back to the macroscopic time scale, thus getting an effective Hamiltonian
completely independent of the scaling and also diagonal in $l$-space.

As mentioned before, the effective Hamiltonian acts on $\bbC^{2}\times\bbC^{2}$-valued
wave functions (including spin); in other words the wave functions
carry a (transversal) angular momentum and a spin index. However, our
effective Hamiltonian is diagonal w.\,r.\,t. angular momentum. Therefore
the reduced space 
$U_{0}L^{2}\left(\Omega_{1},\bbC^{2}\right)=L^{2}\left(\bbR,\bbC^{2}\times\bbC^{2}\right)$
decomposes into two orthogonal sub-spaces, the eigenspaces of $-i\partial_{\vphi}$.
In the following we will take advantage of this by fixing a certain
initial value for $l$ ($1$ or $-1$ in our case) since no transition
between the values can occur in our approximation, i.\,e. we fix
$l$ in Eq.~\eqref{eq:U0def} thereby getting an effective Hamiltonian
depending on $l$ as a parameter. Note that time reversal symmetry transforms
this Hamiltonian into its $-l$ counterpart. This means that a certain
choice of $l$, which is equivalent to restricting the Hamiltonian
to the corresponding eigenspace, violates time-reversal symmetry. 

We stress that this compact result can only be obtained with the revised initial conditions 
introduced above. Had we instead used an initial state with kinetic energies of order $1$ 
(which was our first choice in Section \ref{subsec:Kinetic-term}) 
not only would the first term in Eq.~\eqref{eq:Hfirstorder} contribute to
$H_{\textrm{corr}}$, but also the terms off-diagonal in $l$-space would no longer be
small. As a result, we could no longer switch to macroscopic times by simply dividing by $\veps^2$,
because the first term in \eqref{eq:Hkine} would be of order $\veps^{-2}$
afterwards. Also we could not separate the $l$ and $-l$ states due to the 
off-diagonal terms in $\Heff$. 
While this situation is fully covered by our perturbation
scheme, the resulting effective Hamiltonian would be much more complicated and contain 
different orders in $\veps$ as well as terms allowing transitions from $l$ to $-l$.

\subsection{Structure of the effective hamiltonian}

Although $\Heff$ can be expressed solely by invariants like $\boldsymbol{\sigma}\cdot\boldsymbol{t}$
and geometrical parameters, it is not obvious how the different terms influence spin transport. 
We can make the structure of $\Heff$ more transparent by applying a unitary 
transformation\cite{Michaeli19}
\begin{equation}
U_{1}\left(s\right)=\exp\left(i\sigma_{z}\frac{\pi s}{\wtilde{R}}\right)=\left(\begin{array}{cc}
e^{i\vartheta\left(s\right)} & 0\\
0 & e^{-i\vartheta\left(s\right)}
\end{array}\right)
\end{equation}
($\vartheta\left(s\right):=\pi s/\wtilde{R}$). 

This will not only allow us to write the Hamiltonian in an even more compact 
way but also yield $s$-independent pre-factors, allowing us to determine 
the electronic band structure using  Fourier transformation. With the identities 
\begin{equation}
U_{1}\left(-i\right)\partder sU_{1}^{\dagger}=-i\left(\partder s-i\frac{\pi}{\wtilde{R}}\sigma_{z}\right)\textrm{,}
\end{equation}
\begin{equation}
U_{1}\left(-\partdern s2\right)U_{1}^{\dagger}=-\left(\partdern s2-2i\frac{\pi}{\wtilde{R}}\sigma_{z}\partder s-\frac{\pi^{2}}{\wtilde{R}^{2}}\right)\textrm{,}
\end{equation}
\begin{equation}
\boldsymbol{\sigma}\cdot\boldsymbol{b}=\frac{b}{\wtilde{R}}S+\frac{2\pi R}{\wtilde{R}}\sigma_{z}\textrm{, }\boldsymbol{\sigma}\cdot\boldsymbol{n}=-iS\sigma_{z}\textrm{, }\boldsymbol{\sigma}\cdot\boldsymbol{t}=-\frac{2\pi R}{\wtilde{R}}S+\frac{b}{\wtilde{R}}\sigma_{z}\mmcom\label{eq:invariants}
\end{equation}
and
\begin{equation}
S=\left(\begin{array}{cc}
0 & ie^{-i2\vartheta}\\
-ie^{i2\vartheta} & 0
\end{array}\right)=-U_{1}^{\dagger}\sigma_{y}U_{1}
\end{equation}
we get
\begin{equation}
\begin{aligned}UH_{\textrm{kin},\textrm{e}}U^{\dagger} & =E_{1,1}-\frac{\hbar^{2}}{2m}\left(\partder s-i\frac{\pi}{\wtilde{R}}\sigma_{z}\right)^{2}\\
 & =E_{1,1}+\frac{\hbar^{2}}{2m}\left(-\partdern s2+\frac{2\pi i}{\wtilde{R}}\sigma_{z}\partder s+\frac{\pi^{2}}{\wtilde{R}^{2}}\right)
\end{aligned}
\end{equation}
as well as
\begin{equation}
\begin{aligned}UH_{\textrm{SOC},\textrm{e}}U^{\dagger} & =\frac{\hbar^{2}}{4m^{2}c^{2}}\Biggl(\Phi^{'}\Biggl(\sigma_{y}\left(-i\frac{b}{\wtilde{R}}\partder s-l\frac{4\pi^{3}R^{2}}{\wtilde{R}^{3}}\right)\\
 & +\sigma_{z}\left(i\frac{2\pi R}{\wtilde{R}}\partder s-lb\frac{8\pi^{2}R^{2}+b^{2}}{2R\wtilde{R}^{3}}\right)+\frac{2\pi^{2}R}{\wtilde{R}^{2}}\Biggr)\\
 & +\Phi^{''}l\left(-\sigma_{y}\frac{\pi R}{\wtilde{R}}-\sigma_{z}\frac{b}{2\wtilde{R}}\right)\Biggr)\mmdot
\end{aligned}
\label{eq:Htrafo}
\end{equation}
We now rewrite $U\Heff U^{\dagger}$ in terms of effective fields: 
\begin{equation}
\begin{aligned}H_{\textrm{t}}=U\Heff U^{\dagger} & =E_{1,1}+-r\partdern s2+v\\
 & +\lambda\left(i2\left(\sigma_{y}A_{y}+\sigma_{z}A_{z}\right)\partder s+\sigma_{y}lB_{y}+\sigma_{z}lB_{z}\right)\mmdot
\end{aligned}
\end{equation}
and redistribute terms to obtain 
\begin{equation}
\begin{aligned}H_{\textrm{t}} & =E_{1,1}+r\left(-\left(\partder s-\frac{i\lambda}{r}\widehat{A}\right)^{2}-\frac{\left|\boldsymbol{A}\right|^{2}}{16m^{2}c^{4}}\right)+v\\
 & +\lambda l\boldsymbol{\sigma}\cdot\boldsymbol{B}\mmdot
\end{aligned}
\label{eq:hamilgauge}
\end{equation}
Here we introduced the parameters
\begin{equation}
v=r\frac{2\pi^{2}b^{2}}{\wtilde{R}^{4}}+\lambda\Phi^{'}\frac{4\pi^{2}R}{\wtilde{R}^{2}}\mmcom
\end{equation}
\begin{equation}
r=\frac{\hbar^{2}}{2m}\mmcom\;\lambda=\frac{\hbar^{2}}{8m^{2}c^{2}}
\end{equation}
and the fields $\widehat{A}=\boldsymbol{\sigma}\cdot\boldsymbol{A}$, $\boldsymbol{A}=\left(A_x,A_y,A_z\right)$ and $\boldsymbol{B}=\left(B_x,B_y,B_z\right)$ with components
\begin{equation}
\begin{aligned} & A_{x}=0\mmcom\;A_{y}=-\Phi^{'}\frac{b}{\wtilde{R}}\mmcom\;A_{z}=\Phi^{'}\frac{2\pi R}{\wtilde{R}}+\frac{r}{\lambda}\frac{\pi}{\wtilde{R}}\mmcom\\
\\
 & B_{x}=0\mmcom\;B_{y}=-\Phi^{''}\frac{2\pi R}{\wtilde{R}}\mmcom\;B_{z}=-\Phi^{'}\frac{b}{R\wtilde{R}}-\Phi^{''}\frac{b}{\wtilde{R}}\mmdot\label{eq:effBfield}
\end{aligned}
\end{equation}

The transformed Hamiltonian consists of a kinetic term including a
non-abelian gauge field $\widehat{A}$ and a Zeeman-like term with an effective magnetic 
field $\boldsymbol{B}$. Since the derivatives $\Phi^{\left(n\right)}\left(R\right)$ 
are position-independent, $\boldsymbol{A}$ and $\boldsymbol{B}$ are
constant vectors. Eq.~\eqref{eq:hamilgauge} resembles the Pauli equation 
but there are important differences: 
the gauge field $\widehat{A}$ is matrix-valued and thus the associated 
symmetry group is non-abelian. Also unlike an actual vector potential and 
associated magnetic field, $\widehat{A}$ and $\boldsymbol{B}$ are not related to each other. 
Notice that the last term in Eq.~\eqref{eq:hamilgauge} 
does not violate time-reversal symmetry, since time inversion 
corresponds to the replacement rules $l\rightarrow -l$ and 
$\boldsymbol{\sigma}\rightarrow -\boldsymbol{\sigma}$. 

\begin{figure}
\begin{centering}
\includegraphics[height=5cm]{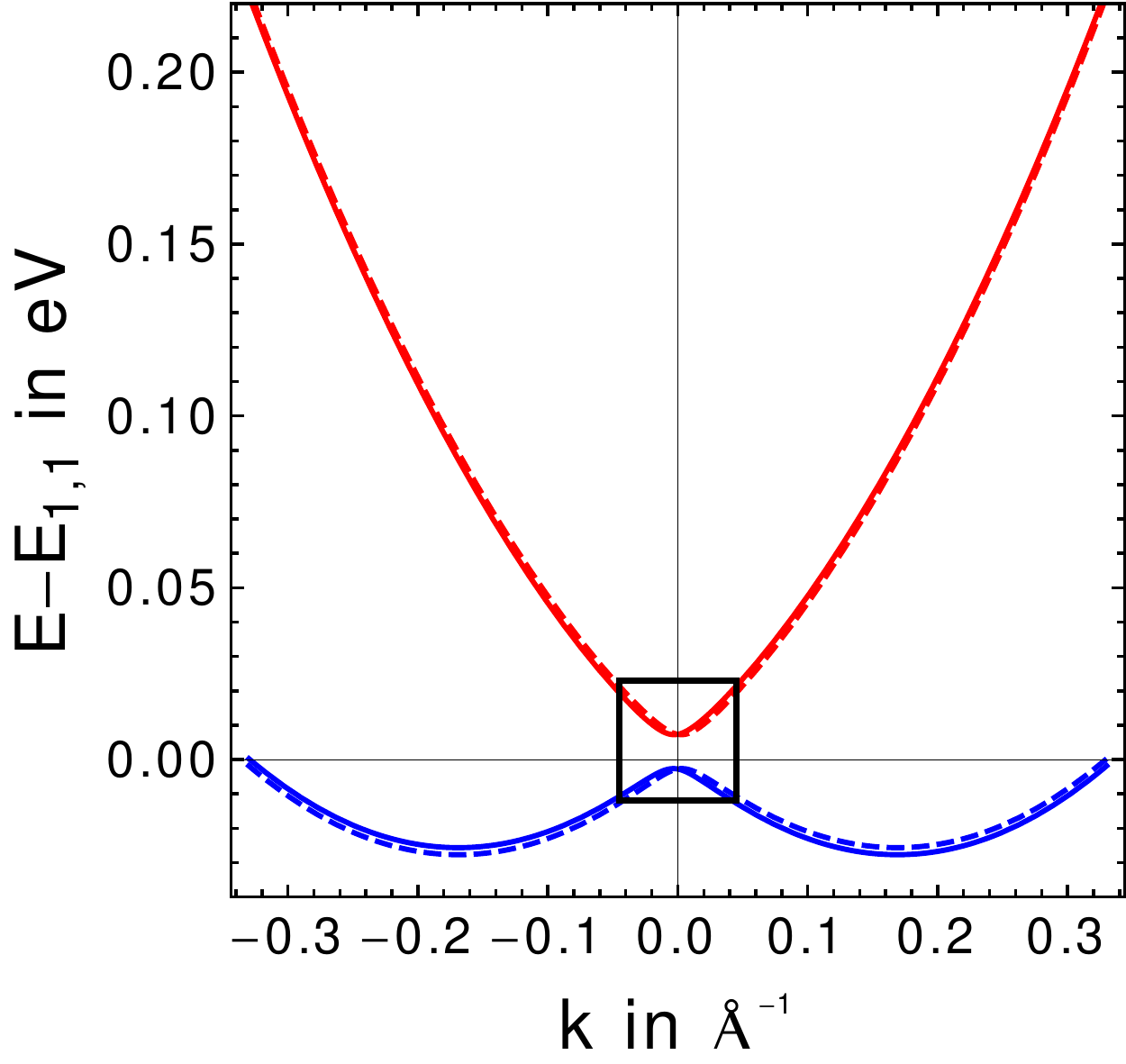}~\includegraphics[height=6cm]{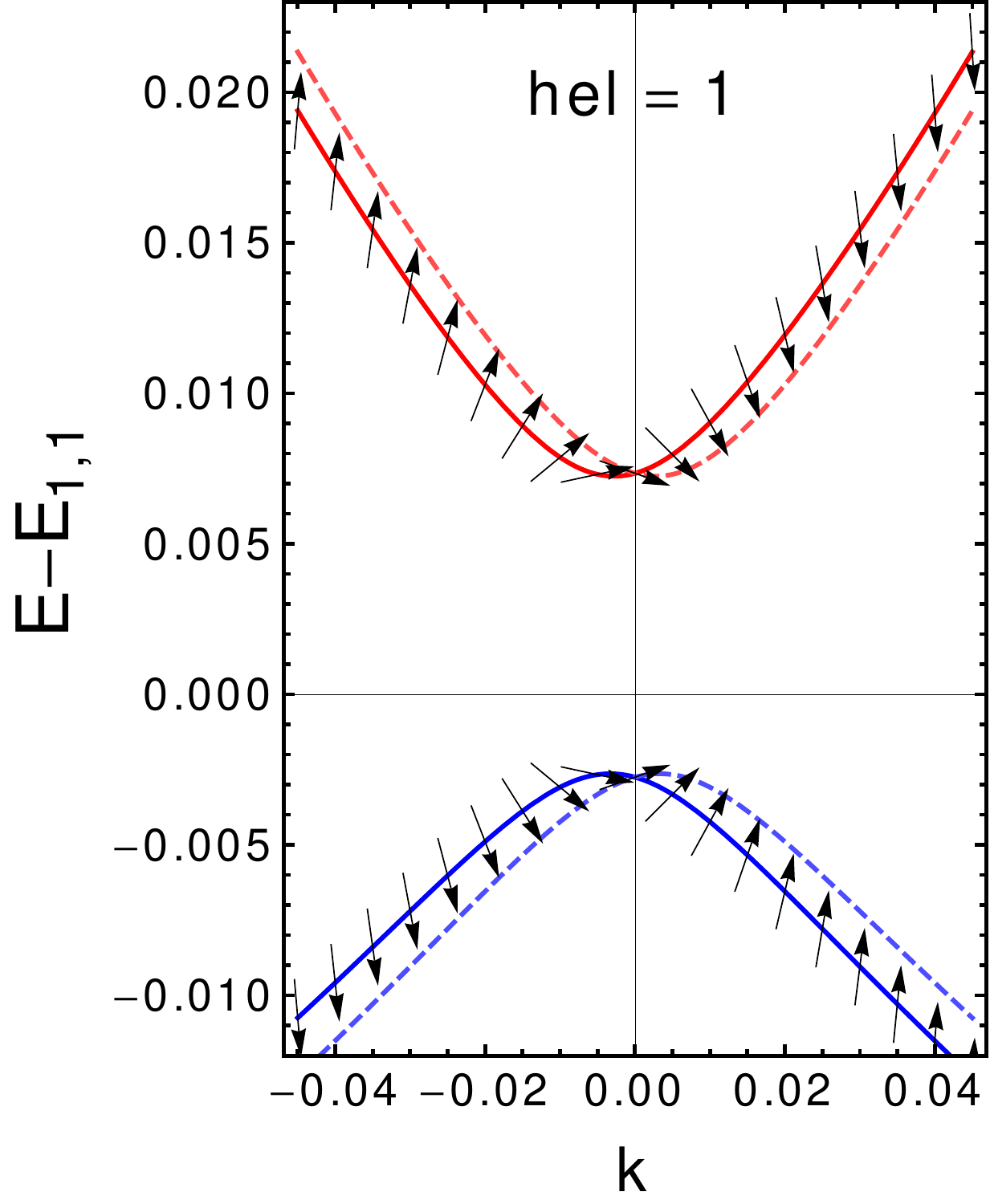}~\includegraphics[height=6cm]{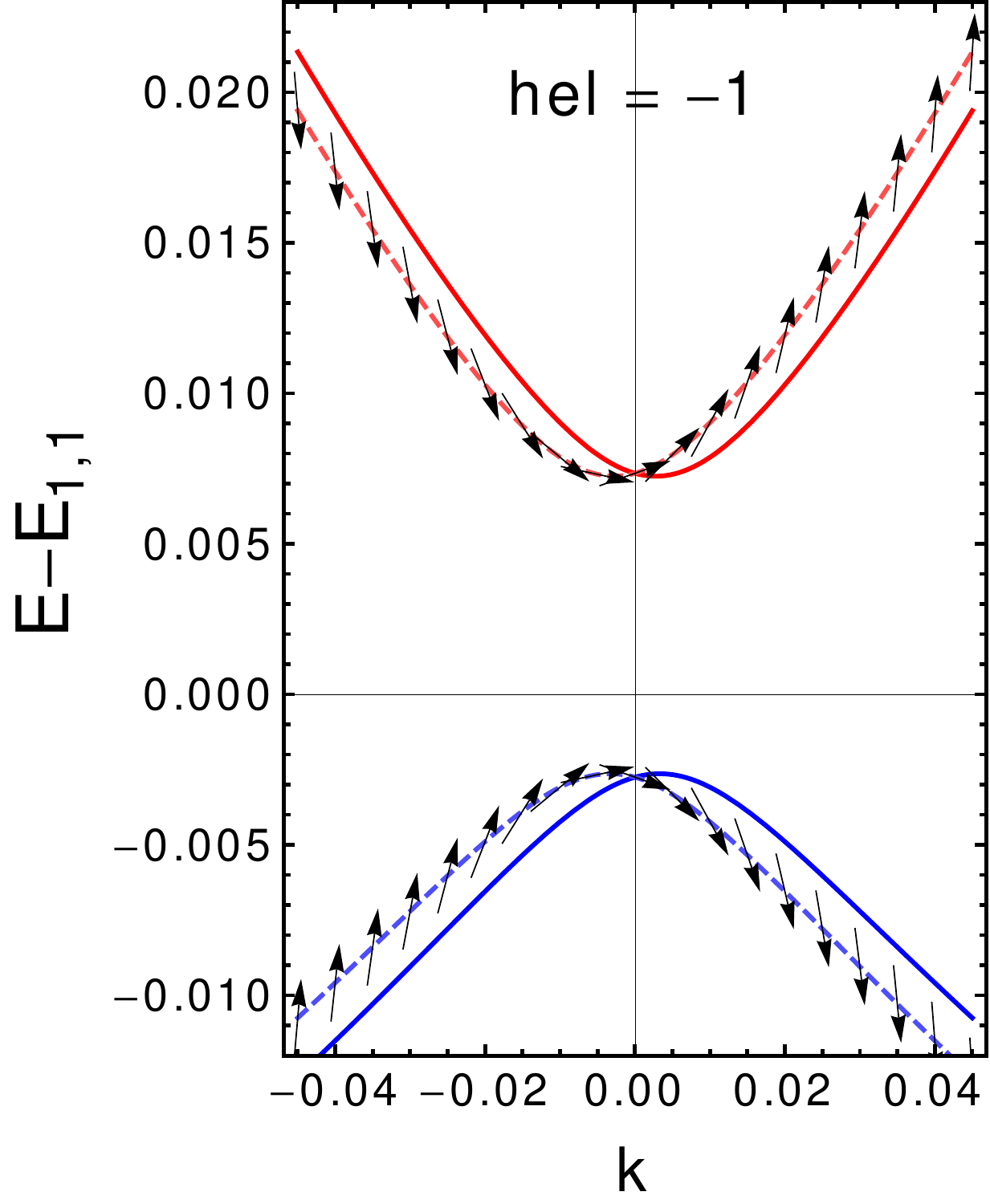}
\par\end{centering}
\caption{Band structure for $R=0.3\,\textrm{nm}$, $b=0.3\,\textrm{nm}$, $r=1\,\textrm{eV}\textrm{Å}^{2}$
and $\Phi^{'}=0.005\cdot\lambda^{-1}\,\textrm{V}/\textrm{nm}$,
$\Phi^{''}=0.005\cdot\lambda^{-1}\,\textrm{V}/\textrm{nm}^{2}$.
The second and third panel show an enhancement of the area around
the gap as well as the spin orientations along the energy band belonging
to $l=1$ for positive ($\textrm{hel}=1$) and $l=-1$ for negative ($\textrm{hel}=-1$)
helicity. The solid lines belong to $l=1$ while the dashed lines
show the $l=-1$ energy band.\label{fig:bandstruc}}
\end{figure}

Taking the Fourier transform of $U\Heff U^\dagger$ we can calulate 
the electronic band structure of $\Heff$ which is shown 
in Fig.~\ref{fig:bandstruc}. The $B_{z}$-component
of the Zeeman term splits the $l=1$ and $l=-1$ energy bands while
the $B_{y}$-component opens a gap of width $\approx2\Phi^{''}\lambda$.
Since $B_{z}$ is small, the splitting is small as well, so the $l=1$
and $-1$ bands lie close to each other. The band structure allows
to analyze some symmetry-related properties of our model. First of
all, changing the helicity means changing the sign of both the pitch
$b$ and the parameter $\wtilde R$ which controls the direction in which the helix
is traversed depending on $s$. However, the angular momentum $l$
always points into the direction of $\boldsymbol{t}$ and thus depends
on helicity itself. This means that for opposite helicities, electrons
with opposite $l$ correspond to one another. Hence, in order to calculate the helicity dependence of spin-related
effects, one has to compare the case with helicity $\textrm{hel}=1$
and $l=1$ to the case with $\textrm{hel}=-1$ and $l=-1$. Therefore
in Fig.~\ref{fig:bandstruc} both helicities are plotted, with the
spin orientation indicated for $l=1$ in the positive and $l=-1$
in the negative case. 

Applying the gauge transformation 
\begin{equation}
U_{\textrm{g}}\left(s\right)=\exp\left(-i\frac{\lambda}{r}s\widehat{A}\right)\label{eq:gaugetrafo}
\end{equation}
leads to the Hamiltonian 
\begin{equation}
\begin{aligned}U_{\textrm{g}}H_{\textrm{t}}U_{\textrm{g}}^{\dagger} & =E_{1,1}+r\left(-\partdern s2-\frac{\lambda^{2}}{4r^{2}}\left|\boldsymbol{A}\right|^{2}\right)+v+\lambda lU_{\textrm{g}}\boldsymbol{\sigma}\cdot\boldsymbol{B}U_{\textrm{g}}^{\dagger}\mmdot\end{aligned}
\label{eq:hamilgaugetrans}
\end{equation}
So if $\boldsymbol{B}=0$, the result \eqref{eq:hamilgaugetrans}
is diagonal in spin space, i.\,e. the spin-orbit coupling can be
removed entirely via the transformation \eqref{eq:gaugetrafo}. This
is consistent with the fact that Rashba-like spin-orbit interaction
terms in a one-dimensional quantum wire can always be removed using
a unitary transformation.\cite{Guo12} Therefore, the Zeeman-like term is
essential for our model to include spin-dependent effects. This term 
stems from the projection of SOC contributions proportional to $\partial_{\vphi}$
with the transversal states, leading to the transversal angular momentum $l$. The occurrence
of the Zeeman term is therefore a direct result of the adiabatic approximation
procedure. It is not present in models based on quantization,\cite{Medina15}
or an approximation that neglects transversal states entirely,\cite{doi:10.1021/jp401705x}
but it appears in Ref.~\cite{Michaeli19} since their effective Hamiltonian
is calculated using rules similar to those we arrived at in Section
\ref{subsec:Adiabatic_approx}. We note, however, that taking into account transversal 
operators as we do is not sufficient to obtain a model with a Zeeman term; it only 
occurs for certain field configurations. Indeed, the Hamiltonian \eqref{eq:hsoczdir} 
coincides with the quantization result and therefore only depends on the intrinsic 
geometry of the helix without retaining any information about the ambient space. 

The Zeeman term is also remarkable since it does not involve momentum operators anymore 
(at least after choosing a certain $l$-eigenstate), 
thus yielding  local (same-site) spin-orbit
interactions after mapping on a discrete tight-binding Hamiltonian, 
as we shall see in the following section. 
Note that the gauge transformation \eqref{eq:gaugetrafo} depends on 
$\lambda/r$ which translates into the spin-orbit to electronic coupling
ratio in the discrete case.

The transformed Hamiltonian also allows us to infer some of the dependence
of spin-related effects on the model parameters. For example, looking
at the expressions \eqref{eq:effBfield} we see that in the case $\Phi^{''}=0$ 
(e.\,g. for a field which is linear in 
the radial coordinate close to the helix radius $R$) 
these effects are very sensitive to the magnitude of the pitch. 
If on the other hand $\Phi^{'}=0$ while $\Phi^{''}\neq 0$ 
(e.\,g. if $\Phi$ has a local minimum at $R$), 
there is no gauge field $\widehat{A}$ present at all and only the $z$-component 
of $\boldsymbol{B}$ depends on the pitch. Even in the limit of a
straight line which can be realized as $b\rightarrow0$ and $R\rightarrow\infty$,
the spin-dependence does not vanish since $B_{y}=-\Phi^{''}\neq0$
remains.

\subsection{The discretized Hamiltonian}

Since we want to employ the Landauer formalism based on Green's function techniques 
to calculate spin transport, we map the continuum Hamiltonian obtained in 
the previous section on a more appropriate discrete tight-binding model.
With the usual rules $\partder s\longrightarrow\frac{1}{2a}\left(\delta_{kj-1}-\delta_{kj+1}\right)$
and $\partdern s2\longrightarrow\frac{1}{a^{2}}\left(\delta_{kj-1}+\delta_{kj+1}-2\delta_{kj}\right)$
we discretize the Hamiltonian:
\begin{equation}
H_{\textrm{t}}=E_{1,1}+-r\partdern s2+v+\lambda\left(i\boldsymbol{\sigma}\cdot\boldsymbol{A}\partder s+l\boldsymbol{\sigma}\cdot\boldsymbol{B}\right)
\end{equation}
from Eq.~\eqref{eq:hamilgauge} to obtain: 
\begin{equation}
\begin{aligned}H_{kj} & \vphantom{\frac{A}{B}}=\left(\varepsilon_{0}+U_{j}\right)\delta_{kj}+t\left(\delta_{kj-1}+\delta_{kj+1}\right)\\
 & \vphantom{\frac{A}{B}}+i\lambda_{1}\boldsymbol{\sigma}\cdot\boldsymbol{A}\left(\delta_{kj-1}-\delta_{kj+1}\right)+\lambda_{2}l\boldsymbol{\sigma}\cdot\boldsymbol{B}\delta_{kj},
\end{aligned}
\label{eq:Hammat}
\end{equation}
with 
\begin{equation}
\varepsilon_{0}=E_{1,1}+\frac{2r}{a^{2}}+v\mmcom\;t=-\frac{r}{a^{2}}\mmcom
\end{equation}
\begin{equation}
\lambda_{1}=\frac{\lambda}{2a}\textrm{, }\lambda_{2}=\lambda\mmdot
\end{equation}
$a$ is the discretization parameter. In second quantization, Eq.
\eqref{eq:Hammat} reads: 
\begin{equation}
\begin{aligned}H & =\left(\varepsilon_{0}+U_{j}\right)\sum_{k}\sum_{\sigma}d_{k,\sigma}^{\dagger}d_{k,\sigma}+t\sum_{k}\left(d_{k,\sigma}^{\dagger}d_{k+1,\sigma}+d_{k+1,\sigma}^{\dagger}d_{k,\sigma}\right)\\
 & +i\lambda_{1}\sum_{k,\sigma,\sigma'}\left(d_{k,\sigma}^{\dagger}\left(\boldsymbol{\sigma}\cdot\boldsymbol{A}\right)_{\sigma\sigma'}d_{k+1,\sigma'}-d_{k+1,\sigma}^{\dagger}\left(\boldsymbol{\sigma}\cdot\boldsymbol{A}\right)_{\sigma\sigma'}d_{k,\sigma'}\right)\\
 & +\lambda_{2}l\sum_{k}d_{k,\sigma}^{\dagger}\left(\boldsymbol{\sigma}\cdot\boldsymbol{B}\right)_{\sigma\sigma'}d_{k,\sigma'}\textrm{.}
\end{aligned}
\label{eq:Hammatsnd}
\end{equation}

Here, $d_{k,\sigma}$ annihilates a particle on lattice site $k$
with spin $\sigma$. It is now apparent that the discretized version
of the Zeeman term $\lambda_{2}l\sum_{k}d_{k,\sigma}^{\dagger}\left(\boldsymbol{\sigma}\cdot\boldsymbol{B}\right)_{\sigma\sigma'}d_{k,\sigma'}$
couples electrons with different spins at the same site as previously
mentioned. Such a term is not present in the hitherto considered phenomenological
methods regardless of  their respective origin with the exception
of Ref.~\cite{Michaeli19}. This brings us to the conclusion that a
carefully performed approximation can lead to substantially different
results. Since we presented a way to derive our approximation scheme
from basic quantum mechanical principles, we are confident that this
additional term is not merely an artifact of the calculation but is
physically justified, provided that the prerequisites for the application
of our procedure are fulfilled in the applications. As the discussion after Eq.~\eqref{eq:hamilgaugetrans} already suggested, this term also
has profound consequences regarding the occurrence of spin-polarization,
which we will further discuss in the results section. 

\section{Transport}\label{sec:Transport}

To investigate the implications of the discretized model from Eq.~\eqref{eq:Hammatsnd}  regarding the CISS effect,
we calculate the transmission and polarization in the Landauer regime. The  quantum mechanical transmission function in a two-terminal setup can be written as\cite{Datta1995,Bell,Paulsson2007}
\begin{equation}
T=\trace\left(\Gamma_{\uptextrm L}\left(G^{\uptextrm r}\right)^{\dagger}\Gamma_{\uptextrm R}G^{\uptextrm r}\right)\label{eq:landauer}
\end{equation}
where
\begin{equation}
G^{\uptextrm r}\left(E\right)=\left(E\mathbf{1}-\Heff-\Sigma\left(E\right)\right)\label{eq:greens}
\end{equation}
is the retarded Green's function\cite{Datta1995} of the isolated molecule 
with the total self energy 
$\Sigma\left(E\right)=\Sigma_{\uptextrm L}\left(E\right)+\Sigma_{\uptextrm R}\left(E\right)$ 
accounting for the coupling to the metallic electrodes and encoding both the strenght of this 
coupling as well as the density of states in the electrodes. The spectral densities 
$\Gamma_{\uptextrm L,\uptextrm R}(E)$ are, as usual, defined in terms of the self-energies as 
$\Gamma_{\uptextrm L,\uptextrm R}(E)=i\left(\Sigma_{\uptextrm L,\uptextrm R}-\Sigma_{\uptextrm L,\uptextrm R}^{\dagger}\right)$. 
To simplify the calculations, we will assume in what follows the wide-band approximation, 
where the self-energies (and hence the spectral densities) are assumed to be purely imaginary, 
energy-independent quantities. We therefore obtain: 
\begin{equation}
\Sigma_{\uptextrm L,\uptextrm R}=-\frac{i}{2}\Gamma_{\uptextrm L,\uptextrm R}\label{eq:wideband1}
\end{equation}
with
\begin{equation}
\left(\Gamma_{\uptextrm L,\uptextrm R}\right)_{ij}\left(E\right)=\sum_{\mu}\wtilde{\gamma}_{\mu}\delta_{i,i_{\mu}}\delta_{i_{\mu},j}\label{eq:wideband2}
\end{equation}
where $\wtilde{\gamma}_{\mu}$ depends on the couplings between lead and molecule and the 
electronic coupling in the lead. So in this case the $\Gamma$`s are just diagonal matrices with 
all diagonal entries either zero or equal to the coupling $\wtilde{\gamma}_{\mu}$ of the lead 
attached to the corresponding site of the molecule.


Assuming that the leads are not spin-polarized, the spin-polarization
can be obtained by calculating the current spin polarization vector\cite{Nikolic2005,Bell}
\begin{equation}
\left(P_{x},P_{y},P_{z}\right)=\mathbf{P}=\frac{\trace\left(\Gamma_{\uptextrm L}\left(G^{\uptextrm r}\right)^{\dagger}\Gamma_{\uptextrm R}\boldsymbol{\sigma}G^{\uptextrm r}\right)}{\trace\left(\Gamma_{\uptextrm L}\left(G^{\uptextrm r}\right)^{\dagger}\Gamma_{\uptextrm R}G^{\uptextrm r}\right)}\label{eq:polarisation}
\end{equation}
where $\boldsymbol{\sigma}=\left(\sigma_{x},\sigma_{y},\sigma_{z}\right)$
is the vector containing the Pauli matrices in the appropriate dimension,
e.\,g. 
\begin{equation}
\sigma_{x}=\left(\begin{array}{cc}
0 & \mathbf{1}_{N\times N}\\
\mathbf{1}_{N\times N} & 0
\end{array}\right)\mmcom
\end{equation}
 The
$z$-component in particular reads
\begin{equation}
P_{z}=\frac{T_{\uparrow}-T_{\downarrow}}{T_{\uparrow}+T_{\downarrow}}\label{eq:P_z}
\end{equation}
with $T_{s}=\trace\left(\Gamma_{\uptextrm L}\left(G^{\uptextrm r}\right)^{\dagger}\Gamma_{\uptextrm R}\pi_{s}G^{\uptextrm r}\right)$
for $s=\uparrow,\downarrow$. 

%
For spin-polarized incoming electrodes we use the transmission function
\begin{equation}
T^\uptextrm{u}=\trace\left(\pi_\uparrow\Gamma_{\uptextrm L}\left(G^{\uptextrm r,\uptextrm{u}}\right)^{\dagger}\Gamma_{\uptextrm R}G^{\uptextrm{r},\uptextrm{u}}\right)\label{eq:landauerpol}
\end{equation}
with $G^{\uptextrm{r},\uptextrm{u}}\left(E\right)=\left(E\mathbf{1}-\Heff-\Sigma^\uptextrm{u}\left(E\right)\right)$ and 
$\Sigma^\uptextrm{u}\left(E\right)=\pi_\uparrow\Sigma_{\uptextrm L}\left(E\right)+
\Sigma_{\uptextrm R}\left(E\right)$ 
which describes a system with its left electrode totally polarized in the spin-up direction
and the analogously defined transmission $T^\uptextrm{d}$ for the spin-down case.

While $T_\uparrow$ and $T_\downarrow$ as used in Eq.~\eqref{eq:P_z} are just 
components of the full transmission for certain spin-channels, the transmissions 
$T^\uptextrm{u}$ and $T^\uptextrm{d}$ that appear in and after Eq.~\eqref{eq:landauerpol} refer to 
transmissions with one incoming spin-channel entirely disconnected from the molecule. 
We define the spin polarization for this scenario by
\begin{equation}
P=\frac{T^{\uptextrm{u}}-T^{\uptextrm{d}}}{T^{\uptextrm{u}}+T^{\uptextrm{d}}}\mmdot\label{eq:Ppolincoming}
\end{equation}
Despite this equation looking almost the same as \eqref{eq:P_z} it is not the expectation value of 
an observable of our model (like the $z$-component of the spin polarization vector 
as in Eq.~\eqref{eq:P_z}) but rather the normed difference of 
transmission functions for two copies of the system with different incoming lead 
configurations (totally up or down polarized). 
The quantity \eqref{eq:Ppolincoming} is closely related to the polarization measured in 
transport experiments with polarized 
electrodes.\cite{Mondal15,Kiran16,doi:10.1063/1.4966237,SMLL:SMLL201602519}

\section{Results}

The transmission and polarization for the Hamiltonian \eqref{eq:Hammatsnd}
were calculated using two sets of geometrical parameters: one for DNA and one
for Helicene as listed in Table \ref{tab:Parameters}. Since we only have one characteristic 
electronic coupling $t$, we chose it to be 0.2 eV for DNA and of the order of 1 eV for helicene. 
The first value is roughly of the order of magnitude obtained in semi-empirical calculations of 
realistic DNA structures,\cite{doi:10.1021/jp801486d} the second value is roughly half of the 
typical $\pi-\pi$ interaction in carbon-based systems. The value of the spin-orbit interaction 
was taken as 5 meV in both cases to have a commnon reference point. It is of the same order of 
magnitude as in our recent estimations for helicene based on a coarse-grained 
model.~\cite{Geyer2019}  The specific values of these parameters have only a quantitative effect 
on our results. Since we have a single electronic state per site on the helix, the obtained 
electronic structure would display no gap; hence, we add a staggering potential 
$U_{k}=(-1)^{k}\Delta\veps$ (with $\Delta\veps=t/2$) to open a band gap mimicking 
the HOMO-LUMO gap in a molecule.

\begin{figure}
\begin{centering}
\includegraphics[scale=0.45]{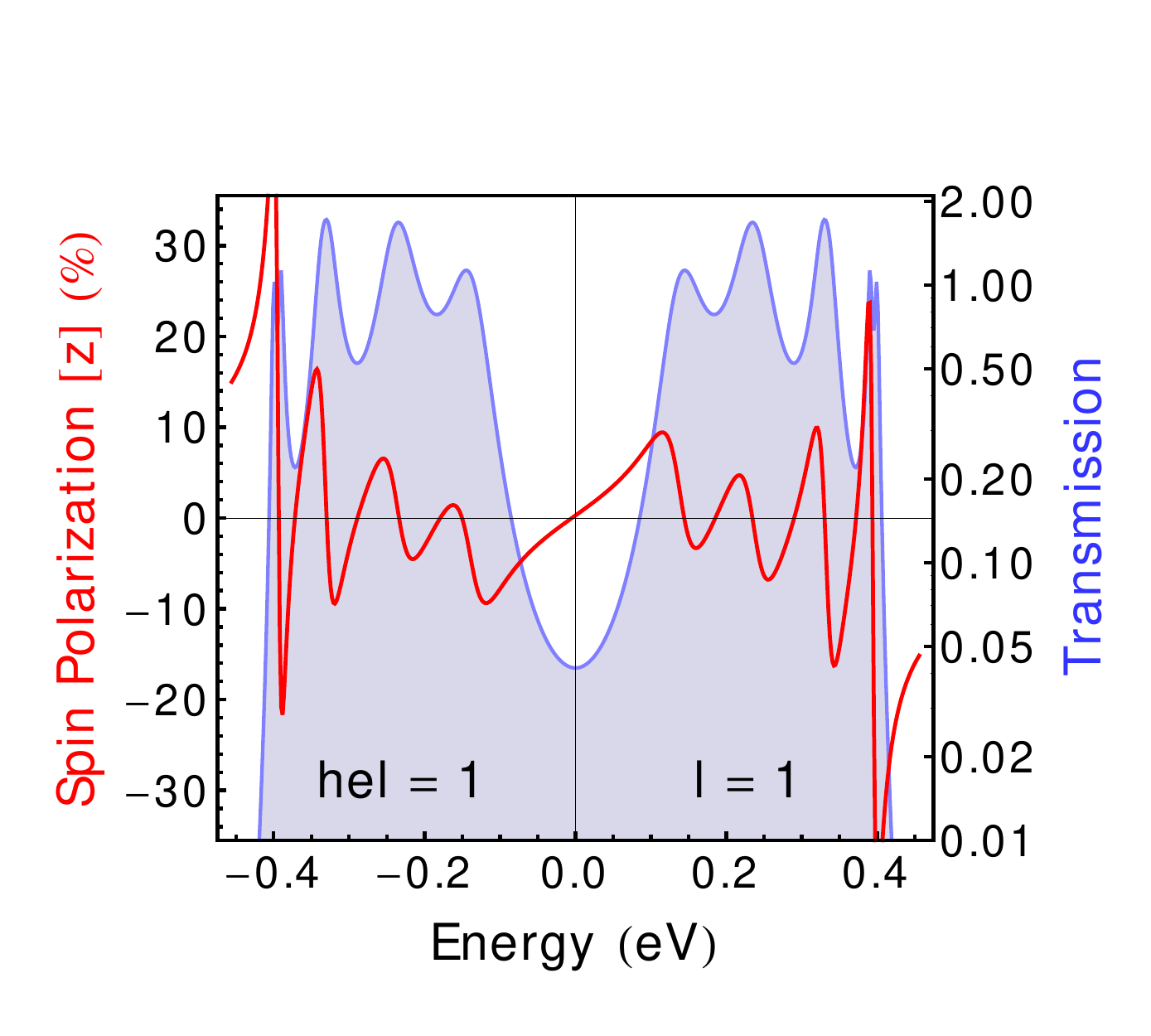}\quad{}\includegraphics[scale=0.45]{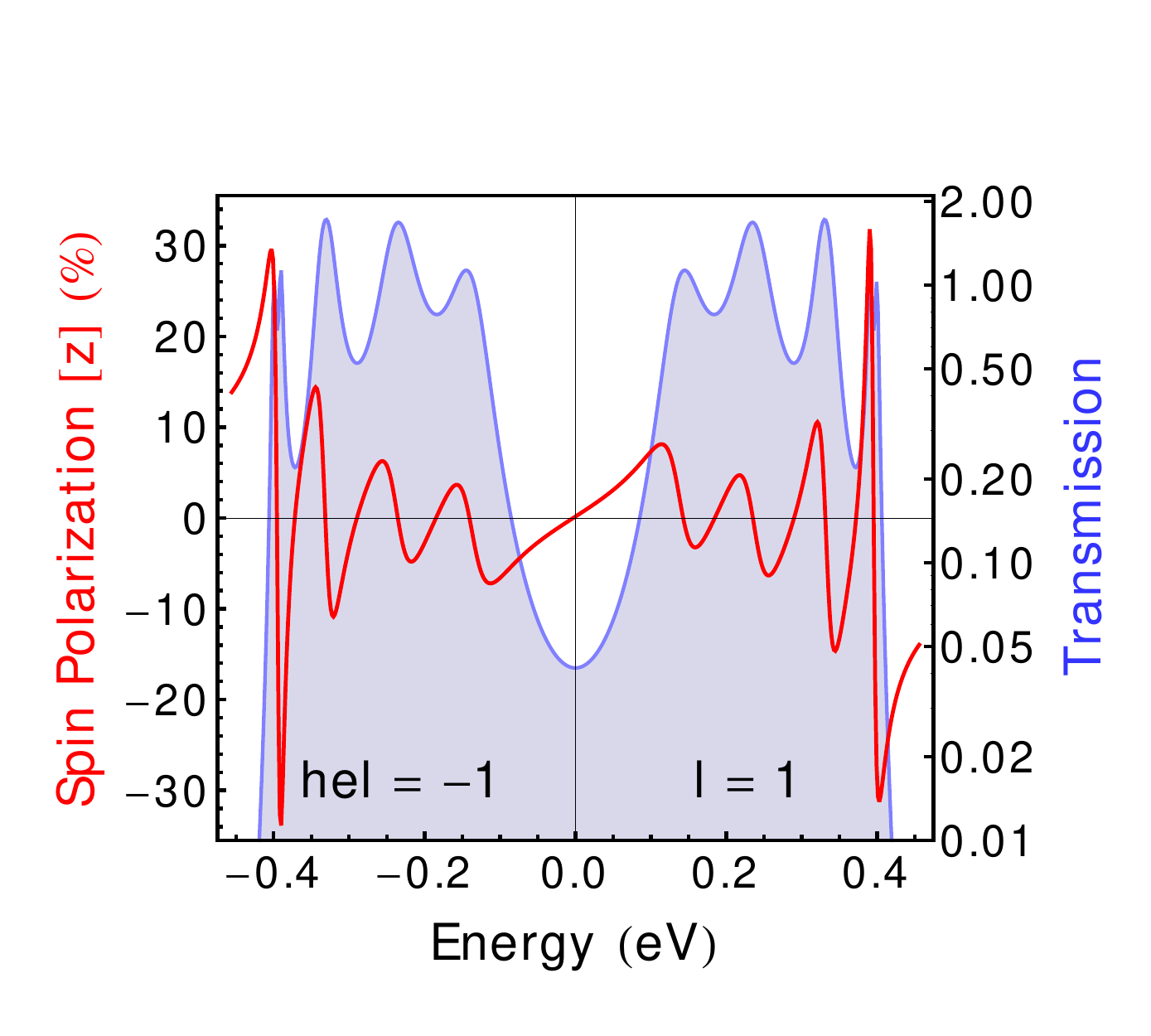}
\par\end{centering}
\begin{centering}
\includegraphics[scale=0.45]{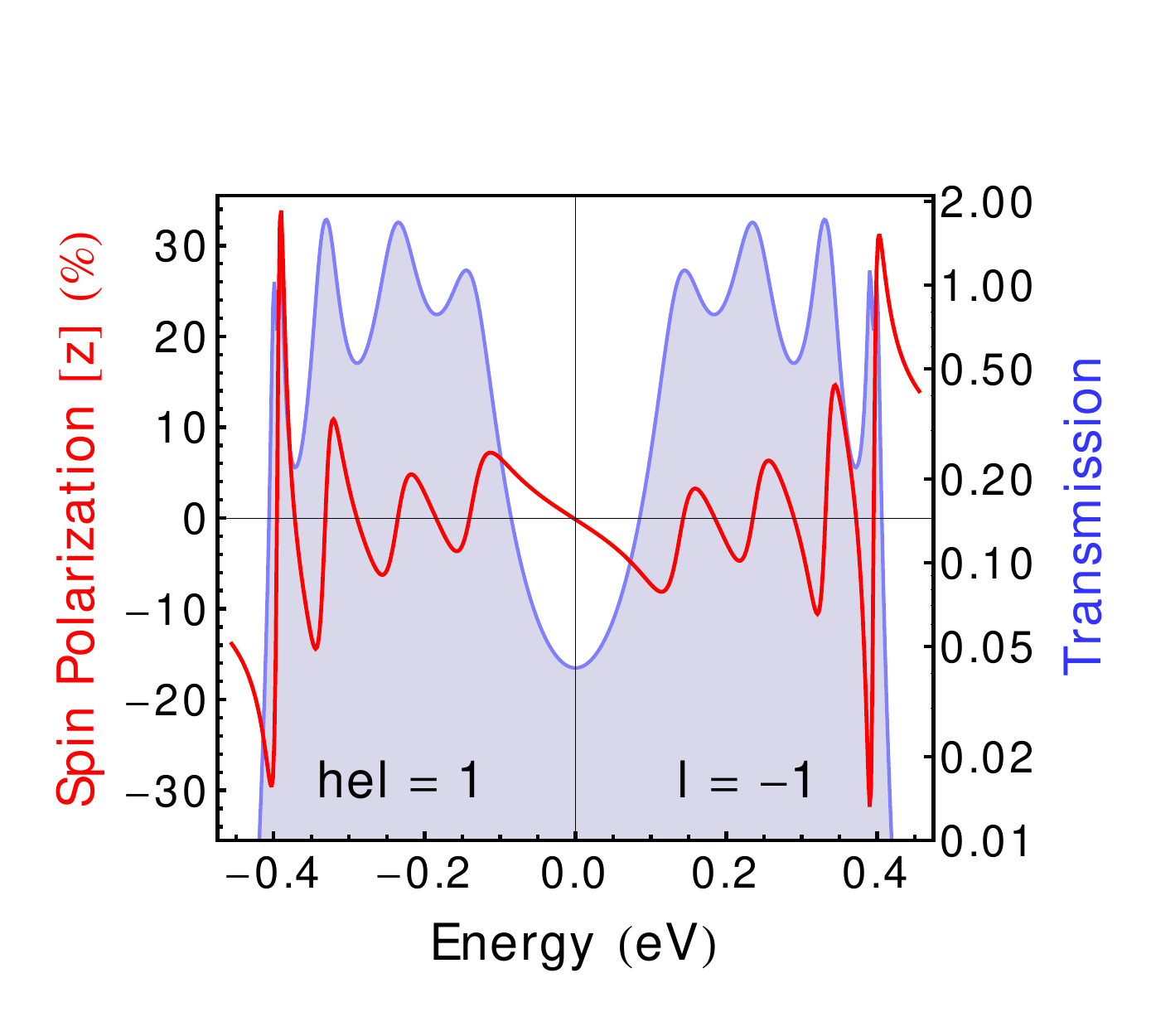}\quad{}\includegraphics[scale=0.45]{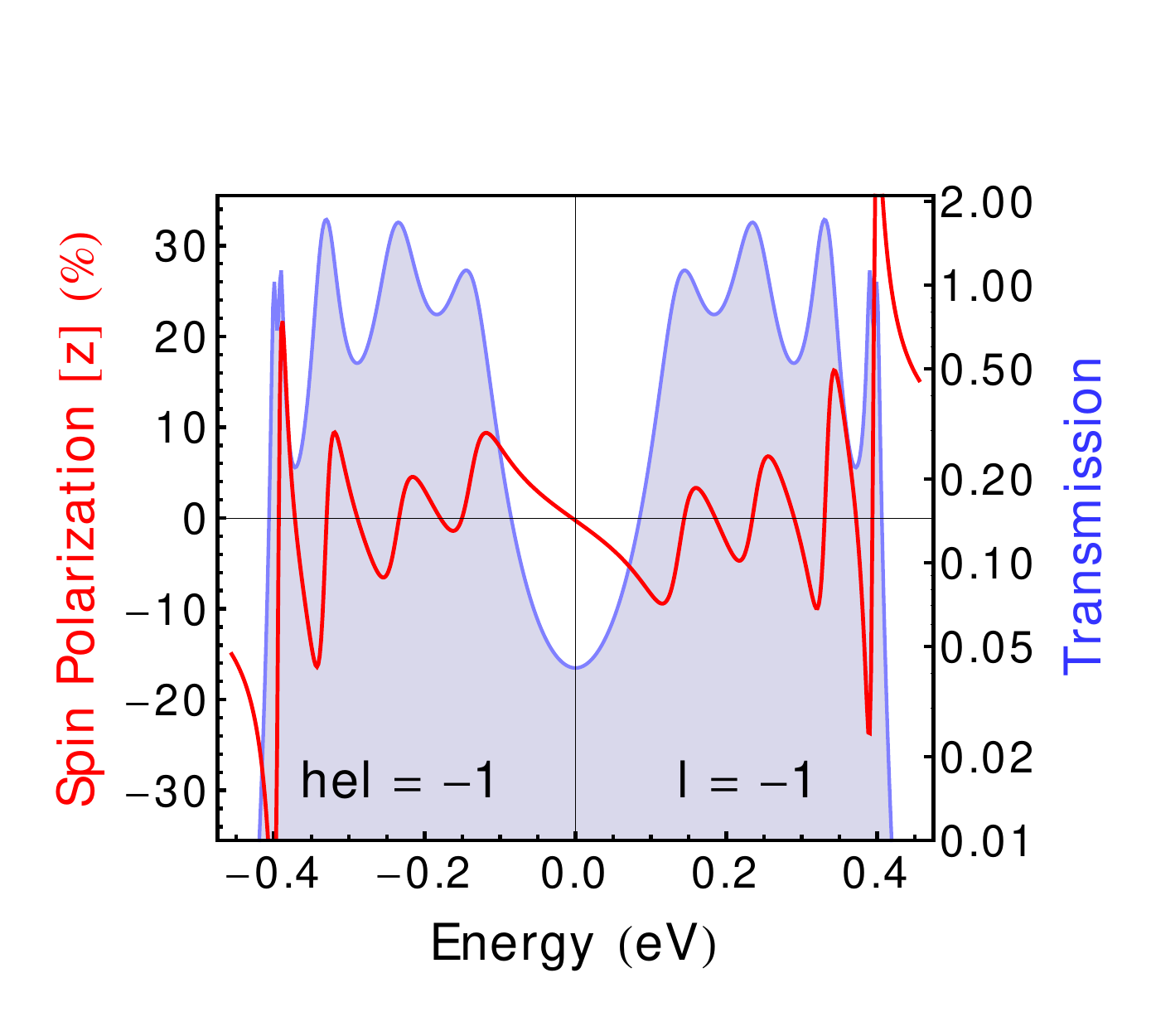}
\par\end{centering}
\caption{Polarization $P_{z}$ (unpolarized incoming leads, Eq.~\eqref{eq:P_z}) 
for different combinations of helicity $hel$
and angular momentum $l$ for DNA. To obtain opposite polarization, 
both the helicity $hel$ of the molecule and the sign of the transversal angular momentum 
$l$ have to be reversed. This is consistent with the fact that the angular momentum points into 
the direction of the tangent vector $\boldsymbol{t}$ and therefore depends 
on the helicity.\label{fig:transpolhell}}
\end{figure}

The results for different combinations of helicity and angular momentum
for DNA are shown in Fig.~\ref{fig:transpolhell}. The polarization
changes sign only if both helicity and angular momentum are reversed,
which is not surprising in light of our statement on the correspondence
of these quantities at the end of Sec.~\ref{subsec:transversal_states}.
Indeed, Fig.~\ref{fig:transpolhell} shows that the spin polarization
depends on the helicity if the corresponding angular momenta are considered. We note that the linear behavior of the polarization around $E=0$ may be a spurious numerical effect related to calculating the ratio of two small quantities when computing the polarization. 
Since electrons with opposite angular momentum have opposite spin polarization, 
the model can only yield a non-vanishing polarization 
if the two angular momenta occur at unequal rates in the initial state 
because otherwise the two contributions would cancel out.

\begin{figure}
\begin{centering}
\includegraphics[scale=0.55]{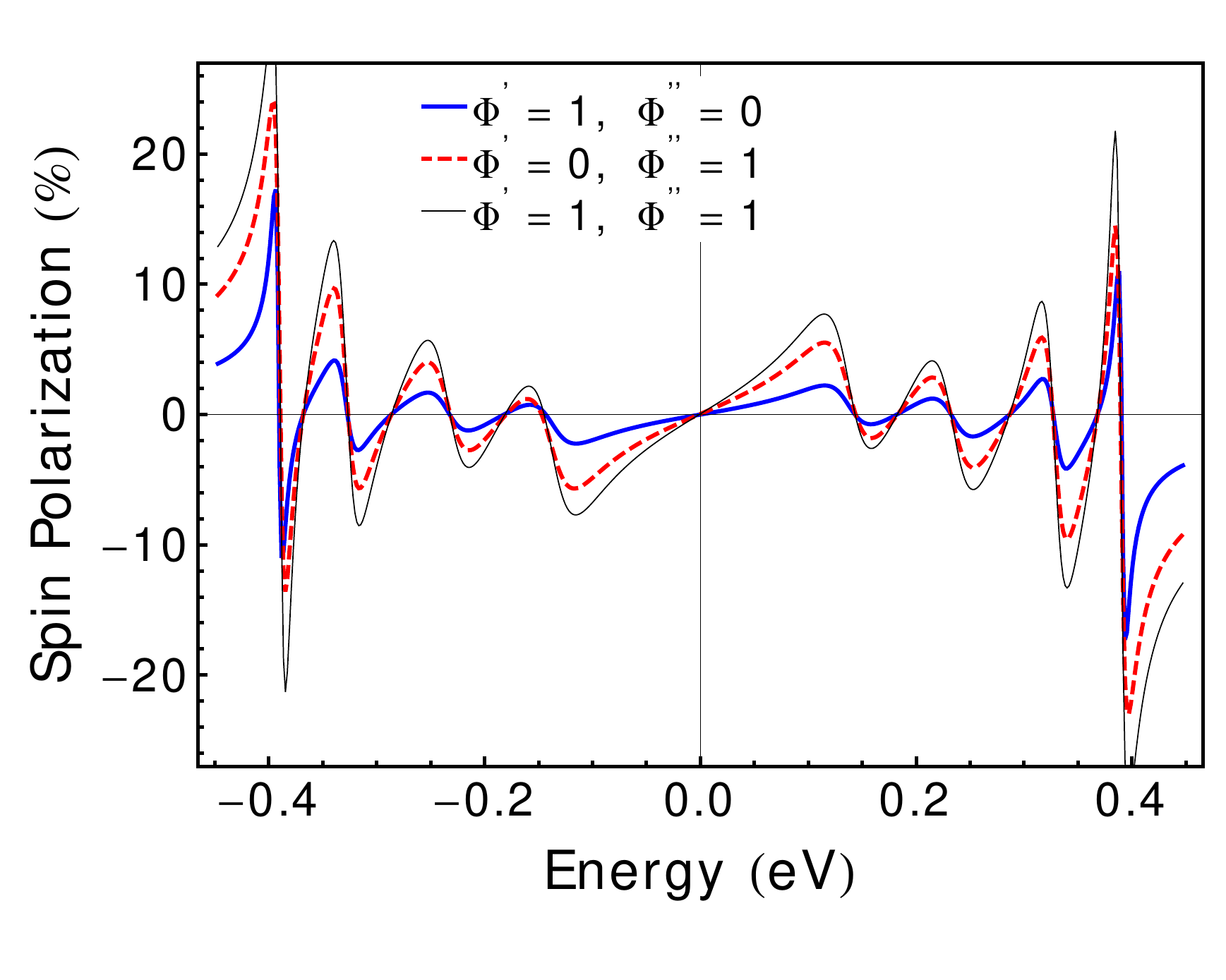}
\par\end{centering}
\caption{Polarization $P_{z}$ (unpolarized incoming leads, Eq.~\eqref{eq:P_z}) 
for different field configurations for DNA parameters.\label{fig:transpolDNA}}
\end{figure}

\begin{figure}
\begin{centering}
\includegraphics[scale=0.55]{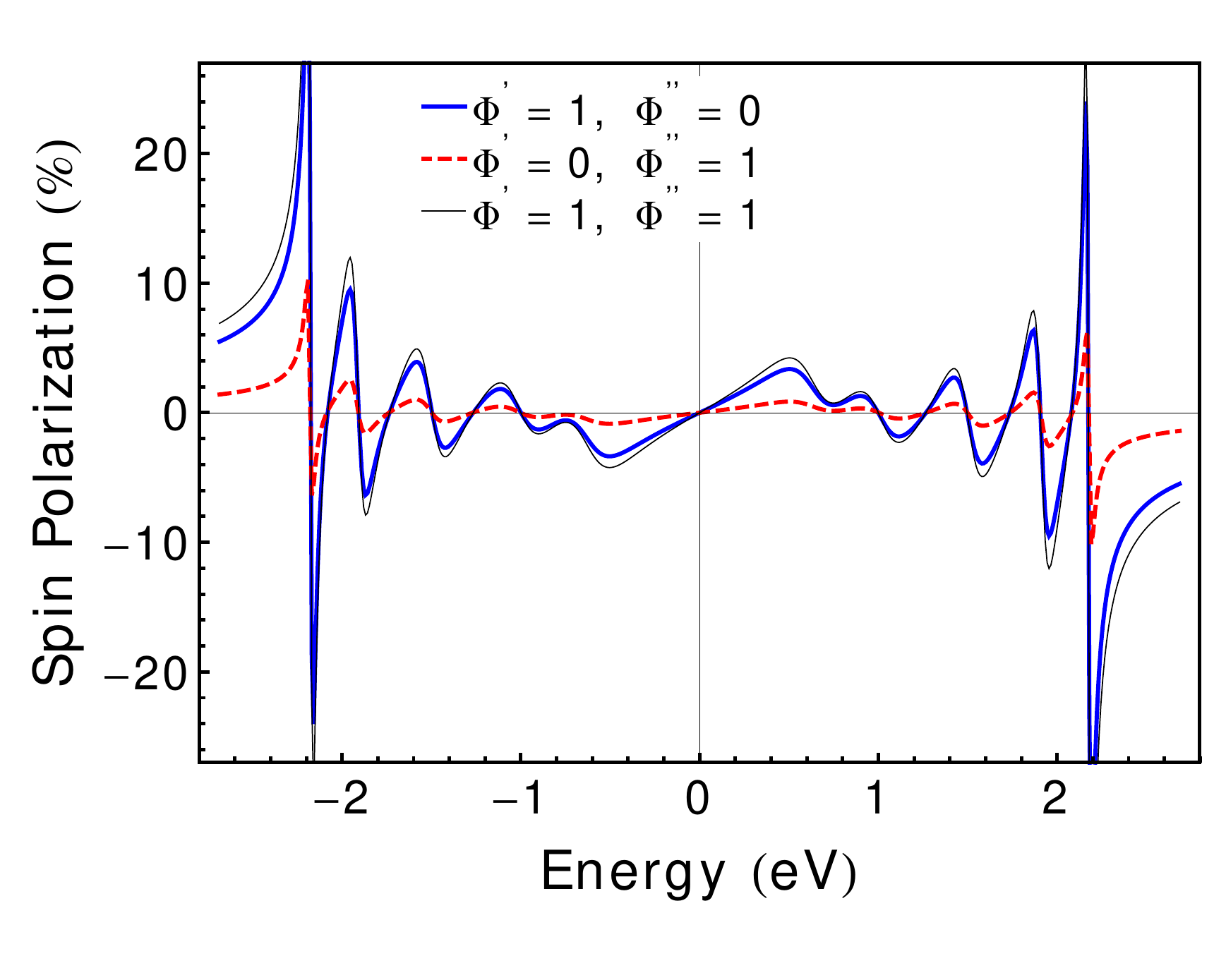}
\par\end{centering}
\caption{Polarization $P_{z}$ (unpolarized incoming leads, Eq.~\eqref{eq:P_z}) 
for different field configurations for helicene parameters. 
Overall polarization is smaller than in the DNA case, but the oscillatory behavior 
is similar.\label{fig:transpolhelicene}}
\end{figure}

In Fig.~\ref{fig:transpolDNA} and Fig.~\ref{fig:transpolhelicene} the energy-dependent spin polarizations for DNA and helicene parameters are shown for different choices of the electric field behavior. 


\begin{table}
\begin{centering}
\begin{tabular}{|c||c|c|}
\hline 
 & DNA & Helicene\tabularnewline
\hline 
\hline 
$t$ (eV) & $0.2$ & $1$\tabularnewline
\hline 
$R$ (nm) & $2.37$ & $0.26$\tabularnewline
\hline 
$b$ (nm) & $3.4$ & $0.36$\tabularnewline
\hline 
$\lambda_{1}\Phi^{'}$ (meV) & $5$ & $5$\tabularnewline
\hline 
$\lambda_{2}\Phi^{''}$ (meV) & $5$ & $5$\tabularnewline
\hline 
$r/\lambda=-t\lambda_{2}/\left(4\lambda_{1}^{2}\right)$ (eV) & $-10$ & $-50$\tabularnewline
\hline 
\end{tabular}\qquad{}%
\begin{tabular}{|c||c|c|}
\hline 
 & DNA & Helicene\tabularnewline
\hline 
\hline 
$A_{y}$ & $-0.445$ & $-0.438$\tabularnewline
\hline 
$A_{z}$ & $-2.163$ & $-189.213$\tabularnewline
\hline 
$B_{y}$ & $-1.366$ & $-4.616$\tabularnewline
\hline 
$B_{z}$ & $-0.406$ & $-1.895$\tabularnewline
\hline 
\end{tabular}
\par\end{centering}
\caption{Parameters used for transport calculations (left), resulting fields
(right, for both $\Phi^{'}\protect\neq0$ and $\Phi^{''}\protect\neq0$).\label{tab:Parameters}}
\end{table}

In qualitative agreement with experimental results,~\cite{Goehler11,Kiran16} the DNA model yields a substantially higher overall spin polarization.
This is probably due to its favorable electronic to spin-orbit coupling
ratio which enhances the contribution from the Zeeman term (see Eq.~\eqref{eq:hamilgaugetrans} and the discussion thereafter). The oscillatory energy dependence is however similar for both systems. 

\begin{figure}
\begin{centering}
\includegraphics[scale=0.55]{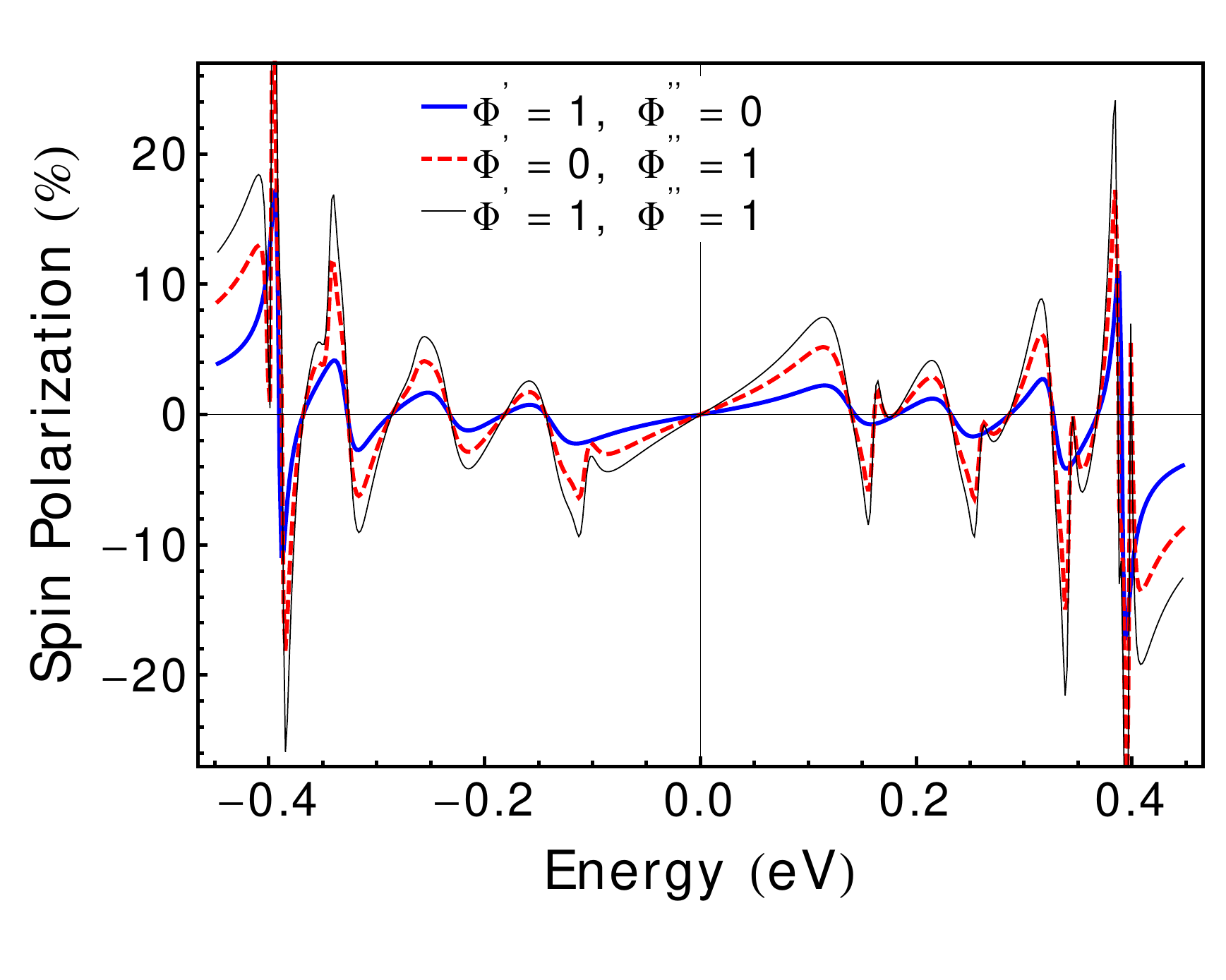}
\par\end{centering}
\caption{Polarization $P$ for polarized incoming leads (Eq.~\eqref{eq:Ppolincoming}) for different field configurations for DNA. Compared to the unpolarized case, 
the polarization is slightly enhanced in some regions 
(sometimes in the opposite direction) but the overall behavior and magnitude 
is similar.\label{fig:oldpolDNA}}
\end{figure}

The polarizations depicted in Figures \ref{fig:transpolhell}, \ref{fig:transpolDNA} and 
\ref{fig:transpolhelicene} were calculated according to Eq.~\eqref{eq:P_z}, i.\,e. for 
unpolarized incoming states. We did the same calculations using Eq.~\eqref{eq:Ppolincoming} 
and found only a negligible difference (less than $0.5\%$) between \eqref{eq:P_z} and 
\eqref{eq:Ppolincoming} for Helicene. For DNA there is a slightly larger deviation; 
yet comparing the result for Eq.~\eqref{eq:Ppolincoming} depicted in Fig.~\ref{fig:oldpolDNA} 
with Fig.~\ref{fig:transpolDNA} shows overall similarity in magnitude and oscillatory behavior. 
This result shows that independently of the spin of the incoming state (which can be controlled 
by selecting the values of the coupling $\Gamma_{L,\uparrow}$ and $\Gamma_{L,\downarrow}$ for 
the corresponding incoming spin channels) the polarization has the same sign for most energies. 

We remark that in this approach a finite spin polarization is obtained in 
the conductance of the system already for a single linear chain with 
a single electronic state per site. This is a consequence of the peculiar 
form of the spin-orbit interaction in Eq.~\eqref{eq:hamilgauge}, 
obtained after projecting the 3d problem on to the effective 1d model. 
This leads to the Zeeman-like term, which cannot be fully removed by 
any unitary transformation as previously discussed.    
Other models required more than one coupled chain with several in-
and outgoing electrodes\cite{doi:10.1021/jp401705x} or, alternatively a decoherence
mechanism described as Büttiker probes\cite{Guo12,Matityahu16} to yield non-vanishing spin 
polarization. Relying on multiple coupled chains to prevent removal of the SOC 
also renders the results sensitive to electronic versus spin-orbit
coupling ratios, which in cases like helicene can lead to very small polarization 
compared to experimental observations\cite{Geyer2019}.

\begin{figure}
\begin{centering}
\includegraphics[scale=0.4]{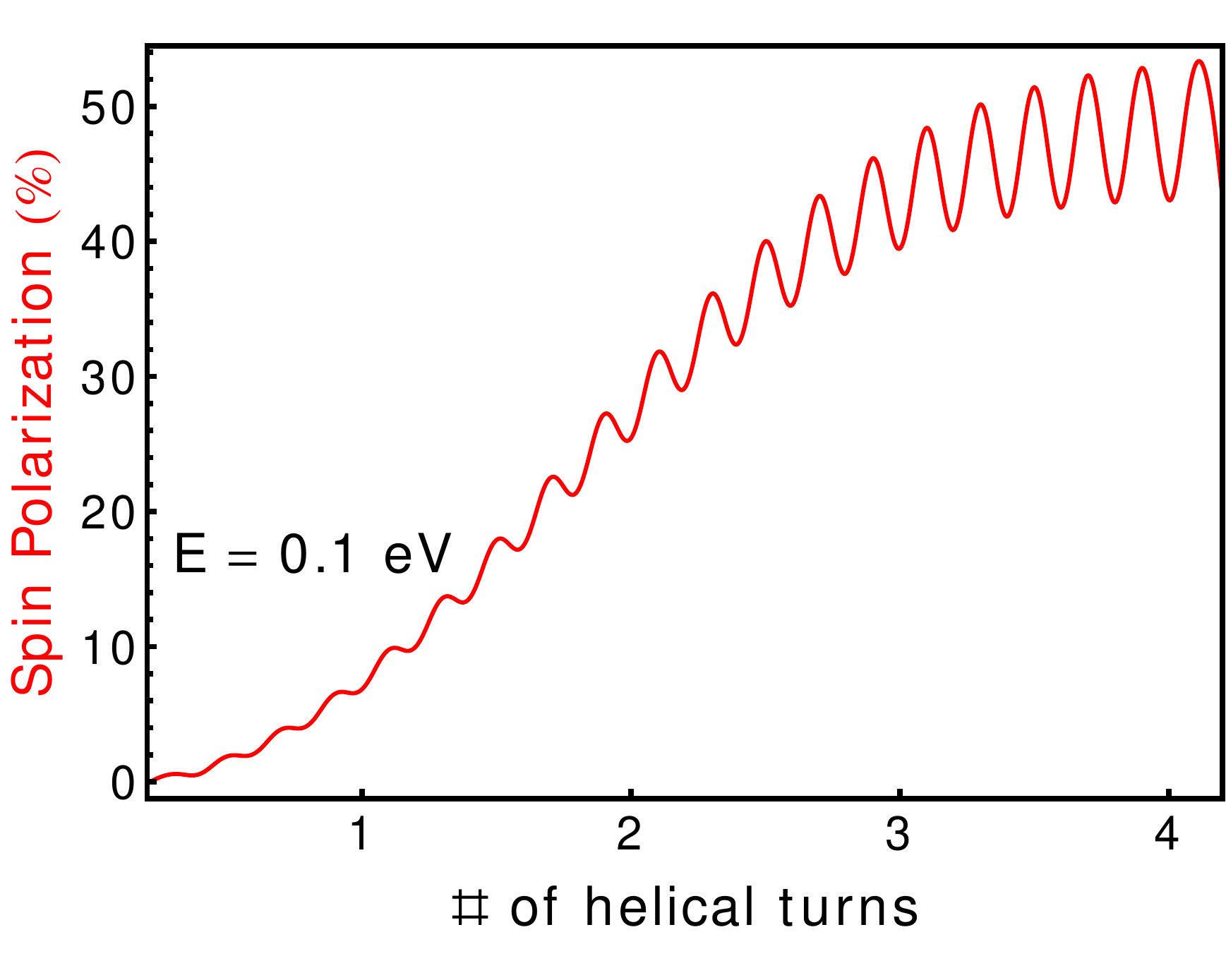}\qquad{}\quad{}\includegraphics[scale=0.4]{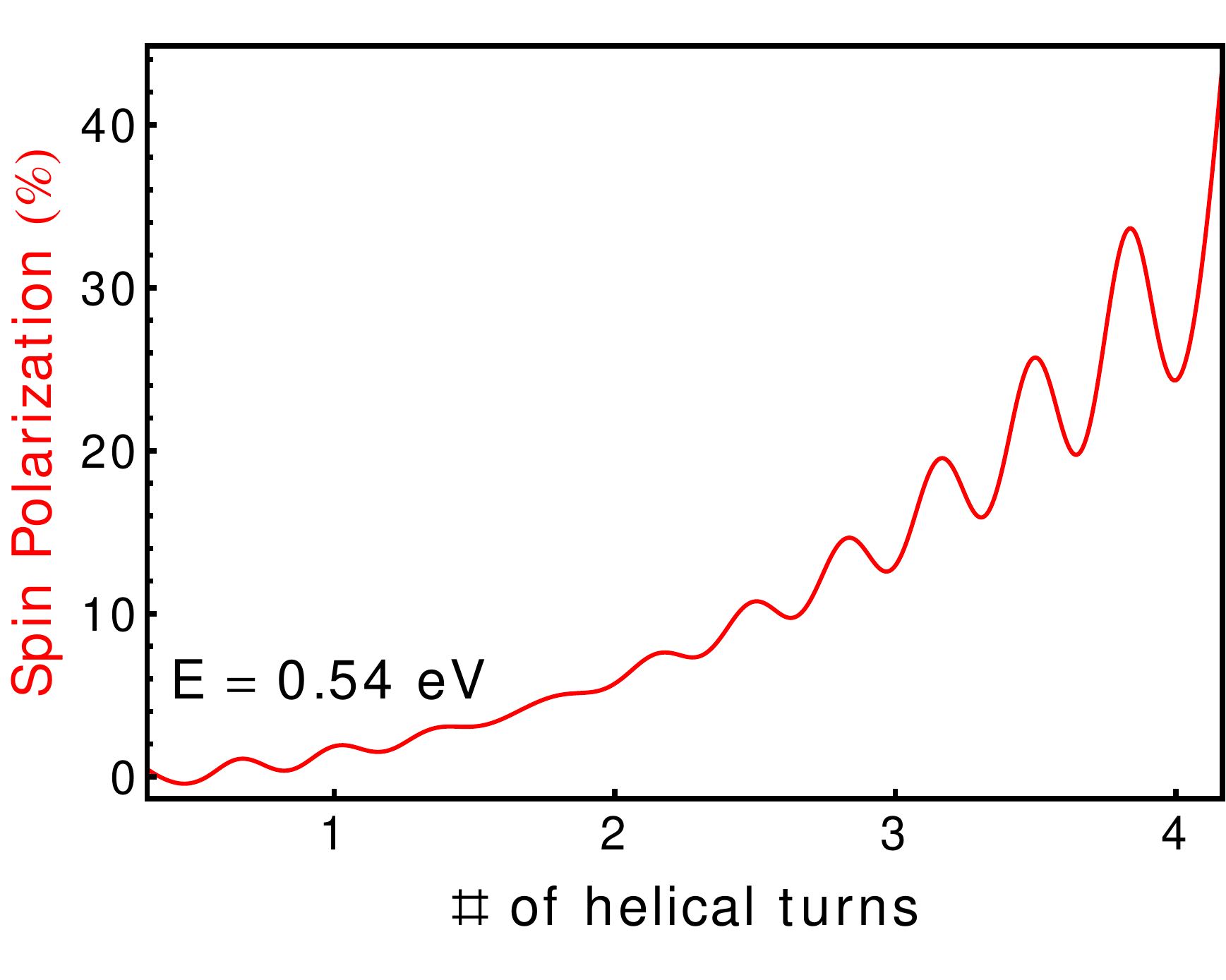}
\par\end{centering}
\caption{Length dependence of the Polarization $P_{z}$ 
(unpolarized incoming leads, Eq.~\eqref{eq:P_z}) for DNA (left) and
helicene at energy $E$ close to the band gap.\label{fig:pollendep}}
\end{figure}

We also investigate the dependence of the spin polarization on the
molecule length. Assuming 10 lattice sites per helical turn for DNA and 6 
for helicene we calculate the polarization at a fixed energy
close to the gap depending on the number of helical turns in the molecule.
The spin polarization increases with the length of the molecule with a roughly 
linear correlation despite oscillations as shown in Fig.~\ref{fig:pollendep} for DNA and helicene.

\section{Conclusions}

We have derived a 1-dimensional effective model for
eletrons moving through a helical confinement. The model includes
 spin-orbit coupling  coming frome a generic scalar field to
describe spin-dependent effects. We have applied a mathematically
sound approximation procedure to show that the effective Hamiltonian
follows from basic quantum mechanical principels in the limit of strong
confinement. Since the adiabatic approximation theory we applied was
developed for a far more general differential geometric setting (but
without spin-dependent effects) it is likely that our approach can
be carried out for a much wider class of systems including spin-dependent
effects as well, starting from confinement to arbitrary curves to
generic submanifolds. Restricting ourselves to the helical case we
obtained a model suited to contribute to the description of the CISS
effect. The model is similar in some respect to the one put forward in Ref.~\cite{Michaeli19}
with which it shares a momentum-independent SOC term that does not
occur in other models, and which is the result of the inclusion of
transversal degrees of freedom during the approximation. This term prevents
removal of the spin-orbit coupling using a gauge transformation as
it is possible in 1-dimensional models lacking the term. One can thus
expect non-zero polarization, as confirmed by our transport
calculations in the linear transport regime. This sets this kind of 
models apart from others that require
multiple incoming and outgoing transport channels or, 
alternatively,  dephasing and decoherence effects to
produce similar results. Being able to obtain similar
terms in tight-binding models might therefore introduce sizeable spin
polarization in a larger class of models.

The Pauli equation~\eqref{eq:pauli} with spin-orbit coupling we started with is 
already an approxi\-ma\-tion of the relativistic Dirac equation for energies which are 
small compared to the rest energy $mc^2$. In this regard our effective Hamiltonian is 
the result of two subsequent limits: the non-relativistic limit and the adiabatic confinement.
These limits are not interchangeable, which was recently pointed out by Shitade and 
Minamitani.\cite{Shitade2020} They showed that by taking the non-relativistic limit of 
the Dirac equation confined to a helix, a spin-orbit coupling of order $\left(mc^2\right)^{-1}$ 
coming from the confinement potential persists, while the usual SOC term in the Pauli equation 
is of order $\left(mc^2\right)^{-2}$. Clarifying the relation between these two approaches 
could help understanding the role of the different SOC terms  contributing to the CISS effect.

\section*{Supplementary Material}

The supplementary information contains a brief summary of the required tools from 
differential geometry, additional details on the adiabatic approximation as well as 
expressions for the spin-orbit coupling terms generated by $s$-dependent fields.

\begin{acknowledgments}
The authors thank Karen Michaeli, Ron Naaman, {{Solmar Varela}}, Vladimiro Mujica, and Arezoo Dianat for very fruitful discussions. This work is funded by the European Union (ERDF) and the Free State
of Saxony via the ESF projects 100231947 and 100339533 (Young Investigators Group
\textquotedblleft Computer Simulations for Materials Design -- CoSiMa\textquotedblright ). 
G.C. acknowledges financial support from the
Volkswagen Stiftung (grant nos. 88366). This work was partly supported by the
German Research Foundation (DFG) within the Cluster of Excellence \textquotedblleft Center for Advancing Electronics Dresden\textquotedblright . We
acknowledge the Center for Information Services and High
Performance Computing (ZIH) at TU Dresden for computational resources.
\end{acknowledgments}

%

\end{document}